\documentclass{aa} 
\usepackage{caption} 

\usepackage{graphicx}
\usepackage{txfonts} 
\usepackage{amsmath}
\usepackage{booktabs}
\usepackage{siunitx}
\usepackage{subcaption}
\usepackage{soul}
\usepackage{natbib}
\bibpunct{(}{)}{;}{a}{}{,}

\begin{document}

\title{The GAGE survey}
\subtitle{I. Deep multi-band stellar variability characterisation and automated classification towards Baade's window}

\titlerunning{GAGE stellar variable candidates}
   \authorrunning{M. Gholami et al.}

\author{
   M.~Gholami\inst{1}\corrauth{mahtab.gholami@inaf.it}
   \and
   M.~Dall'Ora\inst{1}
   \and
   A.~Luna\inst{2,1}
   \and
   J.~Olivares~Carvajal\inst{1}
    \and
   S.~Leccia\inst{1} 
    \and
   M.~Marengo\inst{3} 
   \and 
   G.~Bono\inst{4,5}
   \and
   V.~Braga\inst{5}
   \and
   A.~Calamida\inst{6,1}
    \and 
   G.~De~Somma\inst{1,7} 
   \and
   G.~Fiorentino\inst{5} 
   \and
   M.~Marconi\inst{1} 
   \and 
   D.~Minniti\inst{8,9} 
   \and
   M.~Monelli\inst{10,11,12}
   \and
   I.~Musella\inst{1} 
   \and
   V.~Ripepi\inst{1}
   \and
   P.~B.~Stetson\inst{13}
   \and
   M.~Brescia\inst{14,1}
   \and 
   R.~Bonito\inst{15}
    \and
   M.~Di~Criscienzo\inst{5}
   \and 
   M.~Fabrizio\inst{16,5}
   \and
   M.~Jones\inst{3} 
   \and
   R. Molinaro\inst{1}
   \and 
   S.~Nedhath\inst{17,18} 
   \and 
   C.~Salvo-Guajardo\inst{19}
   \and
   M.~Sánchez-Benavente\inst{11,12}
   \and 
   M.~Tantalo\inst{5}    
}

\institute{
   INAF -- Osservatorio Astronomico di Capodimonte, Salita Moiariello 16, I-80131 Napoli, Italy 
   \and
   European Southern Observatory (ESO), Alonso de Córdova 3107, Vitacura, Casilla 19001, Santiago, Chile 
    \and 
   Department of Physics, Florida State University, 77 Chieftain Way, Tallahassee, FL 32306, USA 
   \and
   Dipartimento di Fisica, Universit`a di Roma Tor Vergata, via della Ricerca Scientifica, I-00133 Roma, Italy
   \and
   INAF - Osservatorio Astronomico di Roma, via Frascati 33, I-00078 Monte Porzio Catone, Italy 
   \and 
   Space Telescope Science Institute, 3700 San Martin Drive, Baltimore, MD 21218, USA 
   \and
   Istituto Nazionale di Fisica Nucleare (INFN) – Sezione di Napoli Complesso Universitario di Monte Sant'Angelo
   Edificio G, Via Cinthia 80126 Napoli, Italy  
   \and 
   Instituto de Astrofísica, Departamento de Fisica y Astronomia, Facultad de Ciencias Exactas, Universidad Andrés Bello, Fernández Concha 700, Las Condes, Santiago, Chile  
   \and 
   Vatican Observatory, Vatican City State, V-00120, Italy 
   \and 
   INAF - Osservatorio Astronomico d'Abruzzo, Via Mentore Maggini s.n.c. 64100 Teramo, Italy 
    \and 
    IAC- Instituto de Astrof\'isica de Canarias, Calle V\'ia Lactea s/n, E-38205 La Laguna, Tenerife, Spain
   \and
    Departmento de Astrof\'isica, Universidad de La Laguna, E-38206 La Laguna, Tenerife, Spain  
    \and 
     Herzberg Astronomy and Astrophysics, NRC, 5071 West Saanich Road, Victoria, British Columbia V9E 2E7, Canada
    \and 
     University of Naples Federico II, Via Cinthia, 80126 Naples, Italy 
    \and 
   INAF - Osservatorio Astronomico di Palermo, Piazza del Parlamento 1, Palermo, Italy, 90134 
   \and
   ASI – Space Science Data Center, via del Politecnico snc, 00133 Roma, Italy 
   \and 
    INAF -- Osservatorio Astrofisico di Arcetri, Largo E. Fermi 5, I-50125 Firenze, Italy  
   \and
   Dipartimento di Fisica e Astronomia, Università di Firenze, Via G. Sansone 1, I-50019 Sesto Fiorentino (Fi), Italy 
   \and 
   Instituto de Astrof\'isica, Pontificia Universidad Cat\'olica de Chile, Av. Vicu\~na Mackenna 4860, 782-0436 Macul, Santiago, Chile 
}

\abstract
{Long-term photometric surveys of the Galactic bulge are vital for characterising its complex stellar populations and structural components, and they provide an essential observational foundation and preparatory framework for next-generation facilities.}
{We present the initial results of the GalActic bulGe with plEiadi (GAGE) project, focusing on the discovery and comprehensive characterisation of variable stars within a field in Baade's window.}
{Utilising archival data from the Dark Energy Camera (DECam) on the Victor M. Blanco 4--m Telescope, we applied high-precision point-spread function (PSF) photometry to a multi-band ($ugriz$) dataset spanning six years. We developed a reproducible multi-step detection pipeline using a generalised Lomb--Scargle analysis to search for periodic signals combined with robust polynomial modelling of the phased light curves to derive accurate mean multi-band magnitudes and luminosity amplitudes.}
{Our selective quality cuts yield a final catalogue made of high-confidence light curves for 6,304 sources. Of these sources, $4,568$ are in common with the Optical Gravitational Lensing Experiment (OGLE\,IV) catalogue, for which we provide a comprehensive multi-band light curve expansion. Additionally, our method uncovered 1,736 ($\sim$ 28\%) variable candidates absent from the OGLE\,IV catalogue. We categorised these sources using a supervised random forest classifier, and notably 1,391 of them were also missing from \textit{Gaia}~DR3 and VISTA Variables in the V\'ia L\'actea (VVV) surveys. This new population is primarily concentrated at the fainter end of our photometric baseline ($i \gtrsim 17.5$~mag), establishing them as new variable candidates enabled by our deeper detection limits.}
{Our results highlight the capability of our photometric pipeline to probe deep, highly crowded stellar environments, and thus it can be used to significantly expand the variable star inventory of the inner Galaxy and serve as a key benchmark for future \textit{Vera C. Rubin} Observatory Legacy Survey of Space and Time (LSST) operations.}


\maketitle 
\nolinenumbers
\section{Introduction} 
The Galactic bulge is one of the most massive and structurally complex stellar components of the Milky Way (MW), and it contains roughly a quarter of the total MW stellar mass within the innermost few kiloparsecs \citep{Valenti10,licquia+15,Valenti2016,Portail2017,Simion2017,cautun2020}. The Galactic bulge stellar populations and kinematics provide critical constraints on early Galaxy formation and evolution \citep{Cao2013, Valenti2016, Portail2017, Simion2017}. Despite decades of study, the relative contributions of early rapid formation processes versus secular evolution through disc instabilities remain debated \citep{Kormendy2004}; however, recent observational and theoretical work has emphasised the role of in situ clumpy star formation at high redshift, suggesting that stellar clumps may contribute significantly to the build-up of the MW bulge mass and structure \citep{Guo2015, Guo2018}. 

Pulsating variable stars are especially powerful probes of Galactic structure and stellar populations. These objects are easily identified thanks to their variability and provide robust distance estimates using well-calibrated period–luminosity (\text{PL}) and period–luminosity–metallicity (\text{PLZ}) relations, enabling three-dimensional mapping of the MW bulge and its constituent populations \citep{Walker1989,Girardi2016,marconi2018,Bono2019,Bono2026}. However, different classes of variables trace distinct structural and evolutionary components. For instance, RR Lyrae (RRL) stars and Type II Cepheids serve as prime tracers of ancient metal-poor populations \citep[e.g.][]{Dekany2015}. In the crowded and highly reddened fields towards the Galactic centre, RRLs are ubiquitous and uniquely map out an unbarred spheroidal structure \citep[e.g.][]{OlivaresCarvajal2024}, distinguishing themselves from other stellar tracers such as red clump stars, which predominantly track younger intermediate-age barred features. Classical Cepheids, on the other hand, are young supergiants that trace active regions of recent star formation and much younger stellar components \citep[e.g.][]{chen2019, Ripepi2021, Trentin2023}. Complementarily, $\delta$ Scuti (DSCT) stars represent intermediate-to-young populations spanning 
the main sequence and subgiant branch. Due to their high spatial density, they serve as 
excellent tracers of fine-scale Galactic disk structure and stellar density gradients 
spanning the solar neighbourhood into the outer bulge components 
\citep[e.g.][]{Poleski2010, Jayasinghe2020}. Consequently, isolating these individual populations provides a cornerstone method for dissecting distinct star formation epochs and structural sub-components that are otherwise difficult to isolate photometrically and spectroscopically \citep{marconi2015, Luongo2024, Muraveva2025}.

The Optical Gravitational Lensing Experiment (OGLE) has been the benchmark for optical time-domain studies in the MW bulge for more than two decades \citep{Udalski2015}. The OGLE\,IV catalogue now contains more than 78,000 RRL stars in the MW bulge and disc, including tens of thousands of fundamental first-overtone and multi-mode pulsators, and it provides extensive time-series $I$--band photometry with exquisite cadence and depth \citep{Soszynski2019} paired with $V$--band data to constrain the stars' positions in the colour--magnitude diagram. The OGLE\,IV surveys have also identified vast samples of other variable types, including DSCT variables, long-period variable (LPV) stars, and eclipsing binary (ECL) stars, further enriching the census of stellar variability in the inner Galaxy \citep{Pietrukowicz2021, Soszynski2013}. 

Synergistic near-infrared surveys, such as the VISTA Variables in the V\'ia L\'actea (VVV; \citealt{Minniti2010}) and its temporal and spatial extension the VVV eXtended (\text{VVVX}) survey \citep{Saito2024}, have pushed variable star studies deeper into highly obscured regions and low Galactic latitudes. By providing multi-epoch infrared light curves, these public surveys effectively mitigate the severe interstellar reddening limitations that hinder optical frameworks, thereby enabling robust three-dimensional structural and kinematic analyses of the inner Galaxy \citep[e.g.][]{Dekany2013,Zoccali2024}.

Despite the transformative impact of OGLE\,IV on the study of stellar variability in the Galactic bulge, its photometric depth and filter coverage impose inherent limitations. The OGLE\,IV survey reaches a typical limiting magnitude of $I \sim 21$ mag in crowded bulge fields, with a peak at $I \sim 18.5$ mag, beyond which completeness rapidly declines. Moreover, generally only a few epochs are available in the $V$--band, making follow-up campaigns necessary, for example for studies in which a multi-band characterisation of the variable stars is needed.

The \textit{Vera C. Rubin} Observatory Legacy Survey of Space and Time (LSST) will provide an unprecedented multi-band and multi-epoch dataset for the southern sky, which will  include a good fraction of the Galactic disc and all of the Galactic bulge. However, early classification of the different types of variable stars is hampered by the adopted observational cadence, and this places some limitations on the use of the LSST data for pulsating star science for the first one to two years of the survey \citep{2023ApJS..265...41D,2024ApJS..273...35D}. Moreover, such a complex and ambitious project would surely benefit from the availability of robust datasets that can be used as benchmark and validation samples of the obtained catalogues in the same passbands. A multi-band light-curve templates are currently missing for the crowded inner Galaxy, despite their necessity for variable classification. 

\begin{figure}[ht!]
    \centering
    \includegraphics[width=\columnwidth]{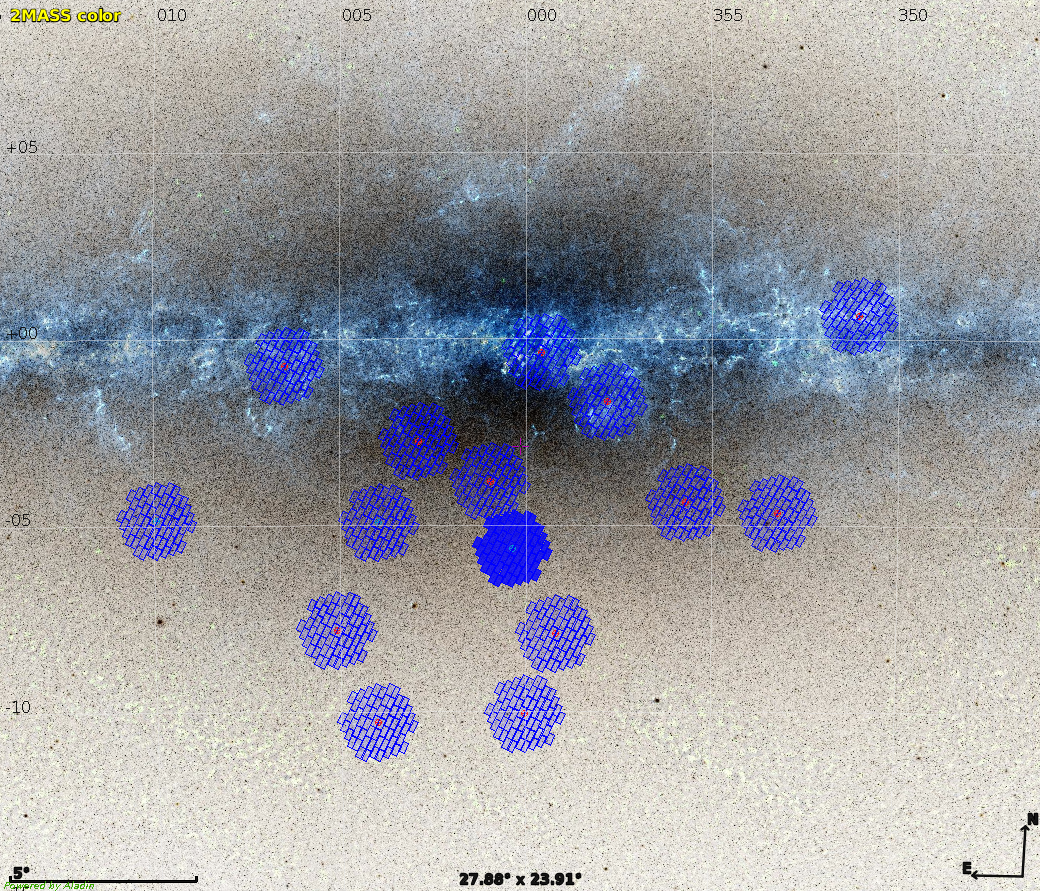}
    \caption{Spatial distribution of the GAGE survey fields (B1 to B15) in Galactic coordinates. The B2 field targeted in this work is highlighted as a filled blue mosaic near the centre ($l \approx 0^{\circ}$, $b \approx -6^{\circ}$). The background consists of a 2MASS colour composite image of the Galactic bulge region.}
    \label{fig:gage_map}
\end{figure}

We have therefore initiated a benchmark project, the GalActic bulGe with plEiadi\footnote{Pleiadi is the high-performance computing facility managed by INAF. Technical information is available at https://pleiadi.readthedocs.io/en/latest/pleiadi/clusters.html.} (GAGE) in anticipation of the upcoming LSST data. The project focuses on the detection and multi-band characterisation of variable stars within the Galactic bulge. The foundational footprint of this survey is built on the fields established by \cite{saha2019}. These observations strategically target crucial structural locations across the Galactic bulge and the forward-directed part of the Galactic bar. These fields are designated as B1 through B6, and the analysis of the first field (B1) was published in the initial study by \cite{saha2019}. Capitalising on the wealth of additional public DECam multi-band time-series datasets available within the Astro Data Lab (NOIRLab) archive that overlap with these inner configurations, we extended this survey by nine additional fields, namely B7 to B15. In Fig.~\ref{fig:gage_map} we show the position of our selected fields against the Galactic bulge.

In this work, we present the detection and characterisation of variable stars in the B2 field, located at $(l, b) = (0\fdg40, -5\fdg70)$ within the low-extinction region of Baade's window. Our methodology combines statistical variability indices, Fourier analysis, robust light-curve fitting, and cross-matching with the OGLE\,IV catalogue. We derive periods, amplitudes, and mean magnitudes across the SDSS $ugriz$ passbands, evaluating the overall completeness and reliability of our detections. Finally, we classify the newly identified variable stars using an optimised random forest machine learning (ML) model. The classification was performed based on multi-band $ugriz$ photometry, utilising the light curve amplitude of each band alongside the calculated Fourier series decomposition coefficients and their corresponding phases as the primary input features. Our DECam-based photometric catalogue reaches approximately up to one magnitude deeper than OGLE\,IV in the $r$--band, substantially extending the parameter space for variability studies in the MW bulge. This increased depth has direct implications for variable identification. By applying a dedicated variability detection pipeline to the DECam time-series data, we identify a large population of variable stars, $\sim$ 28\% of which have not been catalogued previously by OGLE\,IV, demonstrating that a significant fraction of the variable population of the MW bulge remains unexplored at optical wavelengths. These newly identified variables are predominantly fainter and include sources with lower amplitudes or reduced phase coverage in shallower surveys, highlighting the critical role of deep multi-band imaging for achieving a more complete inventory of stellar variability in the Galactic bulge.

This paper is structured as follows: In Sect.~\ref{sec:obs} we describe the observations and data reduction, while Sect.~\ref{sec:method} outlines the methodology and analysis pipeline. Section~\ref{sec:ml_methods} presents the random forest classification framework based on ML. Finally, in Sect.~\ref{sec:results} we discuss the scientific results and distribution properties, with our main conclusions summarised in Sect.~\ref{sec:conclusions}.

\section{Observations and data preparation}\label{sec:obs}

We utilised archival imaging data from the Dark Energy Camera \citep[DECam;][]{Flaugher2015}, mounted on the 4--m Blanco Telescope at the Cerro Tololo Inter-American Observatory (CTIO). The mosaics were prevalently provided by the \cite{saha2019} dataset (acquired under Program ID: 2013A-0719; PI: A. Saha). However, we identified additional archival datasets with a significant overlap of the same spatial pointing, which were subsequently integrated into our analysis to enhance phase coverage. Table~\ref{tab:journal_obs} outlines the number of epochs across all five $ugriz$ filters, alongside the respective project IDs and Principal Investigator (PI) names on which our dataset is based for our selected field (B2) within Baade's window region. As detailed in the table, the considered dataset for this work spans approximately 6 years. The DECam field of view covers approximately $3\,\mathrm{deg}^2$ field of view and has a $0\farcs263\,\mathrm{pixel}^{-1}$ plate scale providing the high spatial resolution required for robust stellar de-blending.

\begin{table}[ht!]
\caption{Observational history of the B2 field.}
\label{tab:journal_obs}
\centering
\setlength{\tabcolsep}{8pt}
\renewcommand{\arraystretch}{1.15} 
\small
\begin{tabular}{llrrrrr}
\hline\hline
\noalign{\smallskip}
Project ID & PI & $u$ & $g$ & $r$ & $i$ & $z$ \\
\noalign{\smallskip}\hline\noalign{\smallskip}
2013A-0327 & Rest           & 0  & 0  & 4   & 0  & 0  \\
2013A-0529 & Rich           & 1  & 1  & 2   & 1  & 2  \\
2013A-0719 & Saha           & 67 & 61 & 86  & 71 & 80 \\
2013A-9999 & Walker         & 0  & 0  & 1   & 0  & 0  \\
2013B-0440 & Nidever        & 1  & 1  & 1   & 1  & 1  \\
2014A-0480 & Rich           & 0  & 1  & 0   & 0  & 1  \\
2014B-0612 & Nataf          & 0  & 1  & 2   & 1  & 0  \\
2015A-0151 & Calamida       & 0  & 0  & 4   & 0  & 0  \\
2016A-0397 & von der Linden & 0  & 4  & 3   & 4  & 0  \\
2016B-0279 & Finkbeiner     & 0  & 3  & 3   & 3  & 2  \\
2018A-0273 & Dawson         & 0  & 19 & 18  & 0  & 0  \\
2019A-0265 & Finkbeiner     & 0  & 3  & 3   & 2  & 2  \\
\noalign{\smallskip}\hline\noalign{\smallskip}
Total      &                & 69 & 94 & 127 & 83 & 88 \\
\noalign{\smallskip}\hline
\end{tabular}
\end{table}

The calibrated and world coordinate system (WCS) corrected mosaics were retrieved from the NOIRLab Astro Data Archive.\footnote{\url{https://archive.noao.edu/}} The dataset consists of 461 individual exposures retained after quality control across the $ugriz$ passbands, with most of the epochs collected (typically around 100, depending on the sky position) in the $r$ band, with exposure times ranging between 30 and 300 seconds. 

The mosaics were split into their individual charge-coupled devices (CCDs), and point-spread function (PSF) photometry was executed on individual single-epoch frames using a dedicated time-series pipeline, based on the \texttt{Daophot\,IV/Allstar/Allframe} package \citep{1987PASP...99..191S, 1990PASP..102..932S, 1994PASP..106..250S} to preserve high-cadence temporal profiles. Image-level quality control cut frames affected by inhomogeneous quality, while catalogue-level filters excluded individual detections with nominal uncertainties $\sigma_{\mathrm{mag}} > 0.1$\,mag. For each target star, individual measurements were matched using \texttt{Daomatch/Daomaster} codes \citep{Stetson93}, in order to build an instrumental light curve.

Since this dataset was composed of archive mosaics coming from different programs, there was a subsequent dithering pattern, which entailed an unavoidable overlap between different CCDs. We took advantage of the overlapping sources, to check the stability of the photometric zero point ($\text{ZP}$) along the field.

Photometric $\text{ZP}$s were derived using \textit{Gaia}~DR3 synthetic photometry, \citep{GaiaSynthetic, Montegriffo2023} obtained by transforming \textit{Gaia} BP/RP mean spectra into the standardised SDSS system \citep{GaiaSynthetic}. Observed instrumental magnitudes ($m_{\rm ins}$) were converted to the calibrated system ($m_{\rm cal}$) for each CCD independently across all $ugriz$ filters using a standard linear relation:

\begin{equation}
    m_{\rm cal} = m_{\rm ins} + \text{ZP} + C \times (c_1 - c_2)
    \label{eq:calibration}
,\end{equation}

\noindent where $C$ represents the colour-term coefficient and $(c_1 - c_2)$ denotes the relevant standard colour index. A comprehensive description of the calibration pipeline, alongside individual transformation equations and empirical coefficients, will be presented in Luna et al. (\text{in prep.}). In the same paper, we will provide a full list of the positions and number of epochs of all our targeted fields from B1 to B15.

We verified that calibrated magnitudes for sources in common between adjacent CCDs usually agreed within $0.02 - 0.03$\,mag in the $rgiz$ bands, while the scatter increases up to $0.1$ mag in the $u$ band. We did not perform a global zero-point correction across the field of view, and we adopt this CCD-to-CCD scatter as a conservative estimate of our internal relative calibration error.

Compiling these single-epoch measurements into a unified master calibration catalogue yielded a static photometric list of $30,415,150$ distinct stellar sources in the B2 field. Each object is tracked via a persistent identifier tied to its coordinate index on a master template frame. In total, the time-series database comprises roughly $9 \times 10^9$ individual photometric detections. Due to the multi-program nature of public archival data and seasonal visibility constraints, this sampling is intrinsically non-homogeneous, meaning that the density of epochs varies by filter and source's coordinates. 

\section{Methodology and analysis variability pipeline} \label{sec:method}

To identify and characterise stellar variability in B2, we developed a reproducible multi-step pipeline. This section details the objects pre-selection, the mathematical framework for their period estimation, and their robust mean magnitudes and amplitudes characterisation, employed in our survey. In the current dataset the $r$ band has the highest temporal resolution (up to 127 independent measurements). This is the reason why it served as our primary baseline for the initial variability search and period determination. Our intimate strategy was to develop a pipeline that almost automatically selects the best promising variable candidates, with a minimum human intervention. This choice naturally excludes several potential candidates, but minimises the need for a check of the individual light curves. This approach will be somewhat unavoidable in an era where LSST will deliver hundreds of thousands of new variable candidates per night.

\subsection{Candidate pre-selection}

Prior to period analysis, each instrumental $r$--band light curve underwent a two-pass robust cleaning procedure to mitigate the impact of photometric outliers. First, every star in our catalogue was analysed excluding the measurements with photometric errors  $\ge 0.1$ mag. Then, a histogram analysis of the remaining measurements helped us in cutting clear outlier measurements. Finally, an initial threshold of $N_{\rm obs} \ge 15$ surviving valid observations was enforced to ensure sufficient phase sampling for the subsequent period search. To isolate high-probability variable stars while excluding saturated sources, we evaluated each object using the Stetson variability index, $L$, computed following Eq.~3 of \cite{stetson1996}'s paper. We adopted a conservative threshold of $L \ge 3$; this minimal baseline was deliberately chosen to ensure that candidate variable stars were retained for subsequent analysis. Sources satisfying this condition were selected for detailed variability analysis, yielding an initial sample of 6,914,581 variable candidates out of the 30,415,150 total initial sources. 

\subsection{Period estimation}\label{subsec:gls}

Primary periods were identified using the generalised Lomb--Scargle (GLS) periodogram \citep{Zechmeister2009}. To optimise the frequency search grid, we dynamically estimated the Nyquist limit for each source based on the average time resolution ($\Delta t$) of the observations, finding the maximum frequency limit of $c_{nyq} = 1 / 2\langle \Delta t\rangle$. Since, on the other side, our data span $\approx 2,000$ days, we considered half of this time window as our upper limit for the period search. Therefore, we conducted the search over a period range spanning from the Nyquist limit up to 1000 days.

The significance of the periodicity was quantified using two primary spectral metrics. First, we identified the \text{Power Peak} ($P_{\rm peak}$), representing the maximum power value in the GLS power spectrum. Second, we defined a Power\_SNR, to evaluate the signal's prominence relative to the spectral noise floor. To ensure uniformity across sources with varying sampling, this metric scales the ratio of the $P_{\rm peak}$ to the root-mean-square error of the entire periodogram ($\sigma_{\rm spectrum}$) by the effective number of observations ($N_{\rm eff}$) remaining after GLS outlier rejection:

\begin{equation}
    \mathrm{Normalised~power\_SNR} = \frac{P_{\rm peak}}{\sigma_{\rm spectrum} \times N_{\rm eff}}
\end{equation}

This normalised ratio accounts for the uneven spatial distribution of individual DECam pointing originating from distinct observational programs. For each CCD, this footprint creates a densely sampled central region exceeding 100 epochs in the $r$ band surrounded by a sparser peripheral area with a secondary coverage peak of fewer than 60 epochs, with a tail down to 20 epochs. Ultimately, 3,820,039 candidates passed this GLS filtering stage. Sources that dropped below the mandatory threshold of $15$ valid observations in the $r$ band after robust outlier mitigation were systematically excluded from the initial candidate sample. 

\begin{figure}[ht!]
    \centering
    \includegraphics[width=\columnwidth]{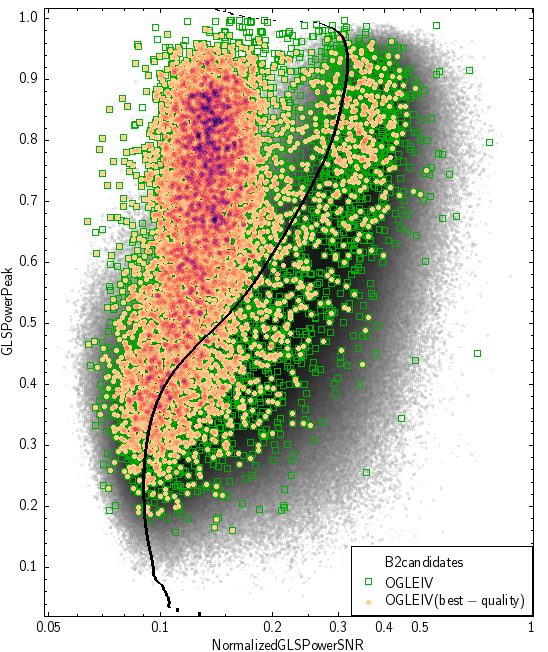}
    \caption{Distribution between GLS power peak vs normalised GLS power\_SNR. Grey points represent the analysed sources, based on our initial selection. Open green squares indicate all the known OGLE\,IV variables in the field. Peach circles highlight the OGLE\,IV variables that with our data show high-quality light curves. The solid black line denotes the threshold used as our final selection boundary, and it marks the 15th percentile distribution in this plane.}
    \label{fig:snr_selection}
\end{figure}

Following these procedures, we analysed the distribution of the sources in the $P_{\rm peak}$ versus power\_SNR, as shown in Fig.~\ref{fig:snr_selection}. The distribution shows two lobes at the high $P_{\rm peak}$ regime. This is due to the normalisation factor $N_{\rm eff}$. The left side of this distribution is populated by candidates with the full number of measurements ($\sim 100$), spatially located in the central part of each DECam CCD. On the other side, the right lobe is typically made by candidates with a significantly smaller number of measurements ($\leq 30$), spatially populating the edges of each CCD. By mapping the distribution of well-characterised OGLE\,IV variables that exhibit the highest-fidelity light curves within our photometry, we established a candidate selection criterion to the 15th percentile of this reference population. This specific threshold was chosen through a joint analysis of the baseline variable distribution and systematic visual inspections of individual light curves. Note that this is a statistical and arbitrary approach, which is focused on pre-selecting the best light curves obtained on the basis of the GLS analysis, and not the intrinsic variable stars distribution. As visible in Fig.~\ref{fig:snr_selection}, this choice excludes a number of potential good light curves, even if located in the right lobe, populated by objects with a limited number of measurements. On the other side, it minimises the number of the poor quality and less sampled light curves that would contaminate our sample and hamper our analysis. In particular, our selection cut $\approx 20\%$ of clean OGLE, IV variables. Therefore, we can estimate an upper limit of $80\%$ for the completeness of the present variable sources analysis. After this phase, we remained with a  sample of 613,341 candidates.

\subsection{Light curve quality validation and robust mean quantities estimation}\label{subsec:ransac}

\begin{figure*}[ht]
\centering
\setlength{\tabcolsep}{2pt} 
\begin{tabular}{cccc}
  \small High-quality & \small High-quality & \small High-quality & \small Marginal (Rejected) \\[2pt]
  \includegraphics[width=0.235\textwidth, height=0.18\textwidth, keepaspectratio=false]{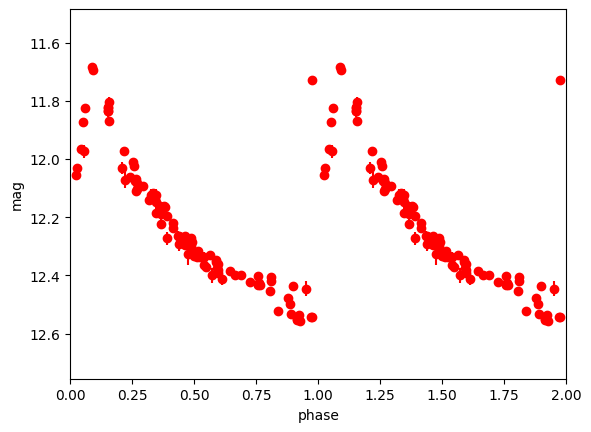} & 
  \includegraphics[width=0.235\textwidth, height=0.18\textwidth, keepaspectratio=false]{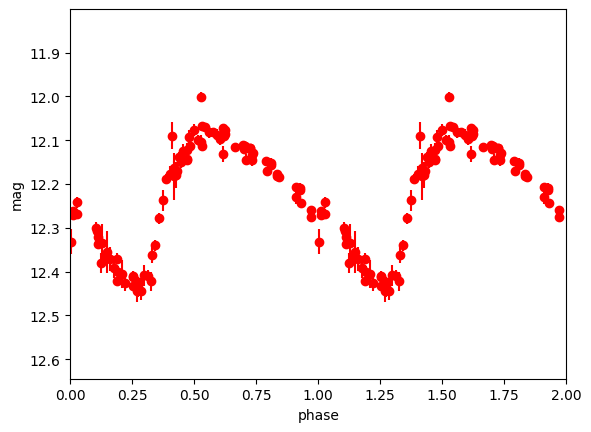} & 
  \includegraphics[width=0.235\textwidth, height=0.18\textwidth, keepaspectratio=false]{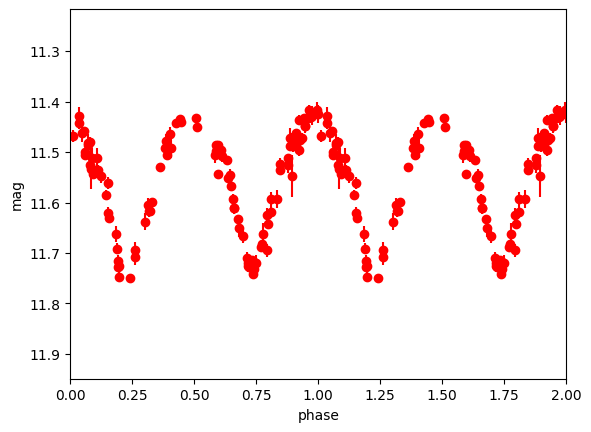} & 
  \includegraphics[width=0.235\textwidth, height=0.18\textwidth, keepaspectratio=false]{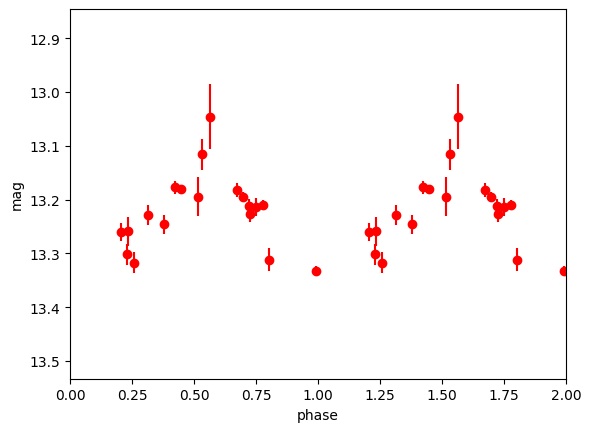} \\[4pt]
  \small (a) New GAGE RRab ($P=0.529$~d) & 
  \small (b) Recovered RRab ($P=0.871$~d) & 
  \small (c) Recovered ECL ($P=0.349$~d) & 
  \small (d) Recovered ECL ($P=2.076$~d) \\
\end{tabular}
\caption{Sample phased light curves in the $r$ band. Panels (a) to (c): High-quality detections identified during visual inspection. Panel (d): Marginal detection of an OGLE\,IV ECL excluded from the final sample due to insufficient phase coverage, as described in the text.}
\label{fig:light_curves}
\end{figure*}

After period identification, to refine the final sample of variable candidates, we implemented a multi-stage validation framework combining automated filtering with visual verification. To efficiently prune this large sample before final fitting, we integrated an automated phase-coverage characterisation into our pipeline. Phased light curves were divided into 10 equidistant phase bins. Candidates were classified as High-Quality (\text{HQ}) if they only contained one empty bin. In contrast, sources with two or more empty bins were flagged as \text{Marginal} due to insufficient phase sampling and were excluded from the final scientific sample. Figures~\ref{fig:light_curves}a--\ref{fig:light_curves}c show a selection of the final HQ light curves with well-defined morphologies, highlighting a newly classified GAGE RRab pulsator alongside recovered OGLE\,IV RRab and ECLs. This refined sample forms the basis of the analysis presented in Sect.~\ref{sec:results}. Additionally, an example of the automated rejection of a light curve is shown in Fig.~\ref{fig:light_curves}d, where an otherwise viable variable is excluded due to sampling gaps.

Following this automated phase-vetting stage and rejecting the alias sources among the candidates, the remaining candidates underwent a thorough, step-by-step visual inspection of their raw light-curve morphologies to confirm the coherence of their pulsational or eclipsing profiles. Following this inspection phase, we used a Random Sample Consensus (RANSAC) regressor \citep{Fischler1981} from Scikit-learn \citep{Pedregosa2011} to derive robust variability amplitudes and clean mean magnitudes for each candidate source. RANSAC is an iterative, non-parametric algorithm designed to fit mathematical models to datasets heavily contaminated by statistical anomalies. Unlike traditional least-squares estimators that are highly sensitive to extreme photometric fluctuations or remnant bad frames, RANSAC randomly samples minimal subsets of the phased data to isolate the true "inlier" population conforming to the underlying light curve structure.

In our method architecture, we utilise RANSAC to fit a high-order polynomial model to each phased light curve. By applying this robust regression framework, photometric outliers are systematically identified and rejected from the final polynomial model fit. For each star, we utilised the resulting optimised inlier distribution to calculate robust, outlier-free, intensity-averaged mean magnitudes and the peak-to-peak variability amplitude ($A$). The algorithm was run 100 times for each light curve, and we took the mean magnitude as the mean in the distribution of individual runs, with the uncertainty being the standard deviation of such a distribution. 

\begin{figure*}[ht!]
  \sidecaption
  \includegraphics[width=12cm]{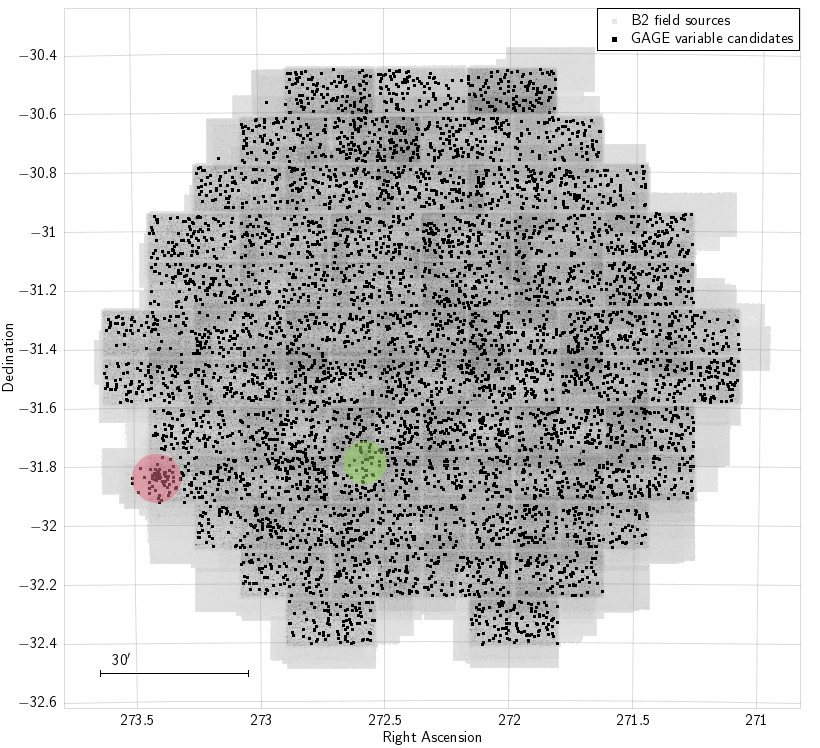}
  \caption{Spatial distribution of the B2 variables. The grey features represent the footprint of the observed field, while the black points highlight the GAGE variable stars detected using our algorithm. A $30'$ scale bar is provided for reference. The distinct overdensity of sources in the lower-left region of the diagram corresponds to variable members belonging to the globular clusters NGC~6558 and NGC~6569, which are highlighted in green and red, respectively.}
  \label{fig:map}
\end{figure*}

A final, rigorous visual check of all light-curve fits was conducted to remove any remaining false detections, resulting in a clean catalogue of 6,304 confirmed variable stars in B2. Of these, 4,568 (72.46\%) are present in OGLE\,IV, while 1,736 are objects genuinely selected by our pipeline. We kept the OGLE\,IV classification for all the further steps, including the OGLE\,IV periods, which are generally based on a larger number of measurements and a wider time span. We show the spatial distribution of our candidate variables in Fig.~\ref{fig:map}, compared to the full catalogue of measured stars. Following visual inspection of the 1,736 newly classified variable candidates, the periods for a subset of objects were adjusted to twice the value determined by the GLS analysis, as their phase-folded light curves clearly exhibited characteristic eclipsing binary profiles. Additionally, We detect a clear excess of variables in the regions around the globular clusters NGC~6558 and NGC~6569 (Fig.~\ref{fig:map}). The variable star populations of these two clusters will be discussed in detail in two forthcoming papers (Olivares et al. \text{in prep.}; Dall'Ora et al. \text{in prep.}, respectively).

It is worth mentioning that candidate stars that did not satisfy our selection criteria and quality-flagging system are not necessarily non-variable objects. In particular, our thresholds explain why we did not recover the entire suite of previously known OGLE\,IV variables within our selected area. For example, in several cases, the valid measurements were below our adopted threshold of $N = 15$. However, the primary objective of this project is not to provide an exhaustive census of all variable stars in our inspected fields in the Galactic bulge, but rather to identify the most robust and highly probable variable candidates.

For the unrecovered or rejected sources, the intrinsic quality of the data may have prevented the GLS algorithm from resolving a clean frequency, or the targets may simply be non-periodic variables. Nonetheless, these objects remain of potential astrophysical interest; their time-domain light curves could support independent studies and constitute a valuable cross-match catalogue for the future LSST alert system. Consequently, the time series of these $\sim 600$\,k stars are available on request. This additional on-request catalogue also includes the OGLE\,IV variables with good light curves that were cut by our selection criteria.

\section{Machine learning classification} \label{sec:ml_methods}

Stellar variability classification for the 1,736 non-OGLE\,IV variables was performed using a supervised ML approach based on the random forest algorithm \citep{Breiman2001}. Given the highly crowded and differentially reddened stellar populations of the inner Galactic bulge, an ensemble decision-tree approach is well-suited to handle non-linear parameter boundaries and minimise cross-classification contamination. This aligns with automated classification architectures implemented in the Galactic centre (e.g. \citealt{Elorrieta2016, Zoccali2024}).

\subsection{Feature space and physical parametrization}

Each light curve was mapped to a 25-dimensional feature vector capturing its key photometric properties. This feature space consists of the primary pulsation period ($P$) derived via the GLS pipeline, multi-band peak-to-peak amplitudes, synchronised colour indices, and localised Fourier decomposition parameters. 

\begin{figure}[ht!]
\centering
\includegraphics[width=\columnwidth]{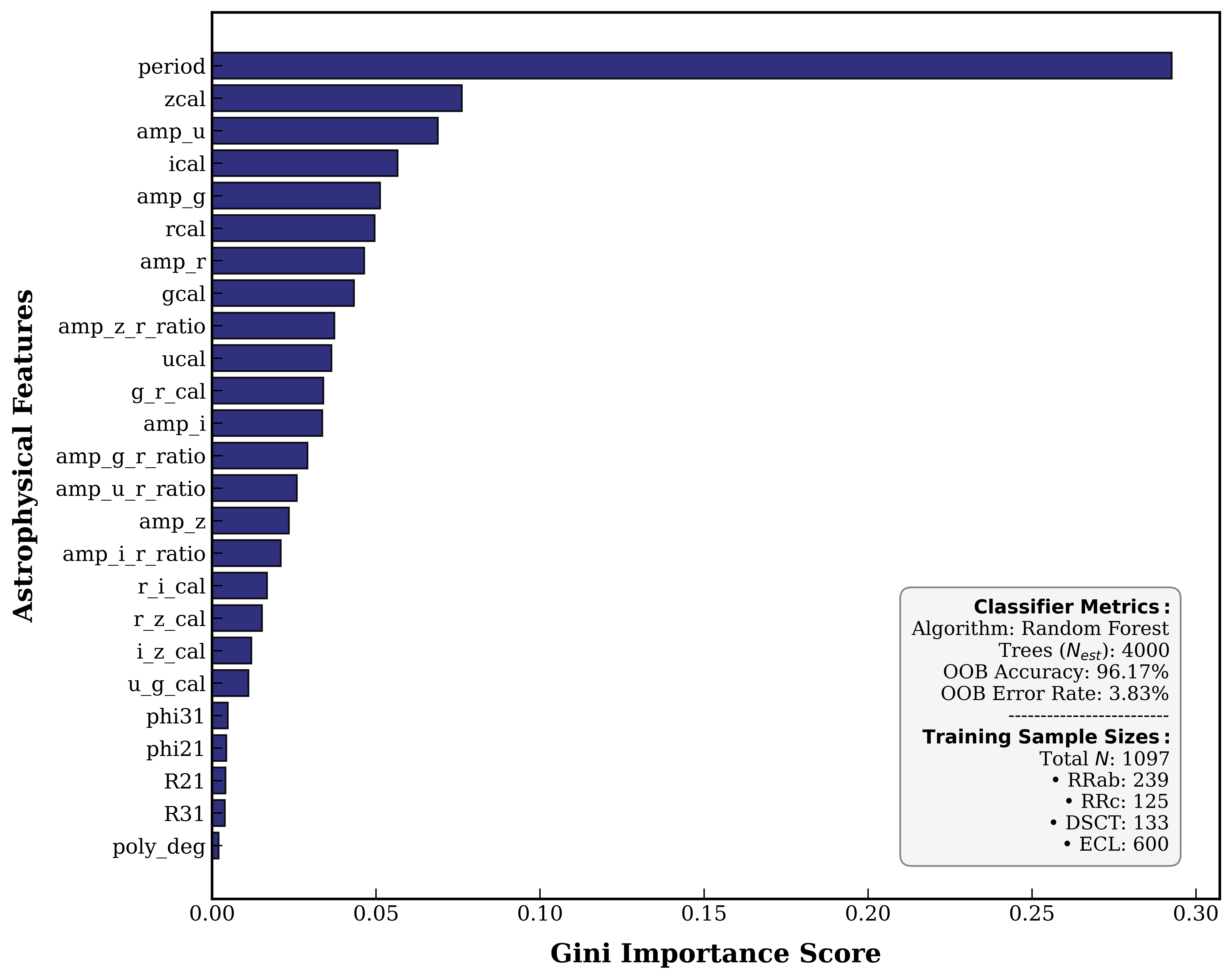}
\caption{Gini importance scores computed from the optimised random forest classifier ($N_{\rm est} = 4000$), ranking the relative leverage of the 25 input features extracted from our multi-band ($u, g, r, i, z$) light curve pipeline. The primary pulsation period dominates the classification with a weight of $\sim$ 29\%, driving the primary separation between major variability classes. Calibrated redder passbands ($z$, $i$, $r$) and $u$--band amplitude ($amp\_u$) form the secondary decision tier to maintain stability against heavy differential extinction. High-order structural parameters, colour indices, and amplitude ratios function as fine-tuned localised phase separators to break degeneracies in narrow boundary regimes. Summary classifier metrics (OOB accuracy = 96.17\%) and training sample sizes ($N = 1,097$, exclusively including ECL, DSCT, RRab, and RRc stars; excluding Cepheids and LPVs) are detailed in the inset box.}
\label{fig:feature_importance}
\end{figure}

To evaluate the relative contribution of each parameter within our random forest architecture, we calculated the Gini importance scores \citep{Breiman1984,Breiman2001} also known as Mean Decrease in Impurity (MDI), directly from the trained classifier (Fig.~\ref{fig:feature_importance}). This metric tracks feature relevance by calculating how much a specific parameter's split points reduce class randomness across all decision trees in the ensemble. 

The primary pulsation period emerges as our dominant classification anchor, accounting for $\sim$29\% of the total model leverage. This is physically expected, as sharp period boundaries cleanly separate different stellar pulsator classes, such as short-period DSCT variables ($P < 0.2$~d) from the core RRL populations ($0.2 < P < 0.9$~d). Beyond the period baseline, the model relies heavily on mean magnitudes in the redder passbands ($z$, $i$, $r$). In highly reddened environments, as the Galactic bulge, these longer wavelengths provide a more stable reference plane for evolutionary positioning than blue filters. 

While explicit colour indices and some amplitude ratios appear lower on the global Gini scale, this does not imply a lack of physical relevance. Because Random Forests optimise non-linear multi-variable splits, the decision trees preferentially exploit the top-ranking raw magnitudes and absolute amplitudes to dynamically reconstruct multidimensional colour spaces and variability scales during training. Conversely, specialised shape parameters such as the RANSAC polynomial degree and higher-order Fourier phase indices also exhibit lower global Gini scores because their impact is localised rather than global. Consequently, these engineered indices, ratios, and structural parameters function primarily as fine-grained classifiers in deeper decision nodes, resolving localised degeneracies to separate asymmetric sawtooth pulsators (such as RRab stars) from symmetric ECLs within closely overlapping parameter regimes.

As displayed in the embedded metrics panel of Fig.~\ref{fig:feature_importance}, the internal evaluation of the forest relies on the out-of-bag (OOB) framework \citep{Breiman2001}. Because each of the $N_{\rm est} = 4000$ decision trees is trained on a bootstrapped subset of the training data, the remaining unselected templates serve as an independent, built-in validation set during model construction. The model achieves an unbiased global OOB accuracy of 96.17\% (an OOB error rate of 3.83\%) on the training matrix. This high performance is maintained despite the significant class imbalance within the overall reference matrix sourced from the recovered OGLE\,IV sources within our detected variable candidates, which is physically dominated by ECLs ($N_{\rm ECL} = 3,940$) relative to the pulsator subpopulations ($N_{\rm RRab} = 299$, $N_{\rm RRc} = 156$, $N_{\rm DSCT} = 166$). To ensure balanced decision boundaries across the ensemble, internal class weights were applied inversely proportional to each group's sample frequency during tree construction.

\subsection{Training strategy and validation performance}

To verify the ML performance prior to classifying our unlabelled variable candidates, the baseline model was trained and validated using a cross-matched reference sample of confirmed variable stars from the OGLE\,IV catalogue. Feature extraction was restricted strictly to 4,568 variables common between OGLE\,IV and our dataset, including the pulsation period, $ugriz$-band amplitudes, colour indices, and Fourier parameters.

Crucially, not all common sources between OGLE\,IV and our variable candidate catalogue were included in the training set. The training framework explicitly focused on four primary target classes: ECL, DSCT, RRab, and RRc. insufficient statistics variable types in our sample, such as classical Cepheids and Type II Cepheids (6 objects in total), and LPVs (zero in our sample) were intentionally excluded from the training set due to their low overall incidence in the cross-matched reference sample, which prevented the formation of a statistically viable training pool. Additionally, one object was omitted due to incomplete multi-band photometry (lacking $ugriz$-band amplitude measurements required for feature extraction), leaving a final curated reference pool of 4,561 sources.

The ECLs numerically dominate the reference pool ($N = 3,940$), which can create a majority-class bias during model optimisation. To prevent algorithmic dominance and preserve sensitivity to rarer pulsator classes, we implemented a stratified sampling strategy. For the pulsator classes, an $80\% / 20\%$ stratified split was applied, allocating $80\%$ of the sources to the training phase (239 RRab, 125 RRc, and 133 DSCT stars, as detailed in Table~\ref{tab:sample_distribution}). For the ECL population, the training subset was capped at a strict limit of $N = 600$ sources to maintain class parity across decision trees to prevent algorithmic dominance.
\begin{table*}[ht!]
\caption{Comprehensive breakdown of the cross-matched OGLE\,IV reference catalogue.}
\label{tab:sample_distribution}
\centering
\begin{tabular}{lcccccc}
\hline\hline
\noalign{\smallskip}
Class & Total pool & Training set & Validation support & Purity & Completeness & F1--score \\
      & (OGLE\,IV)  & (Stratified pool) & (Independent test) & (Precision) & (Recall) & \\
\noalign{\smallskip}\hline\noalign{\smallskip}
RRab  & 299   & 239   & 60    & 0.9062 & 0.9667 & 0.9355 \\
RRc   & 156   & 125   & 31    & 0.8000 & 0.9032 & 0.8485 \\
DSCT  & 166   & 133   & 33    & 1.0000 & 1.0000 & 1.0000 \\
ECL   & 3,940 & 600   & 3,340 & 0.9985 & 0.9961 & 0.9973 \\
\noalign{\smallskip}\hline\noalign{\smallskip}
Total / W. Avg & 4,561 & 1,097 & 3,464 & 0.9951 & 0.9948 & 0.9949 \\
\noalign{\smallskip}\hline
\end{tabular}
\tablefoot{
Dataset distribution alongside validation performance metrics evaluated on features in common with our variable candidates. The F1--score denotes the harmonic mean of model purity (precision) and completeness (recall).
}
\end{table*}
The remaining 20\% of the pulsator populations, together with the remaining unselected ECLs, formed an independent validation set ($N = 3,464$). This separate dataset was used to evaluate our model’s final performance on common variables with OGLE\,IV before applying the trained ensemble to our unlabelled variable candidates.

The final classification for any source is assigned based on the ensemble's output probability vectors. To maximise final catalogue purity, we enforced a conservative dual-threshold selection logic: a candidate is assigned to a pulsator class only if its primary probability vector satisfies $\ge 0.55$, whereas the ECL population requires a more stringent threshold of $\ge 0.85$. These operational limits were calibrated empirically by benchmarking the random forest ensemble against the OGLE\,IV truth tables. We determined these probability boundaries by systematically testing a range of threshold values to find the optimal balance between maximising the recovery of genuine variable stars (completeness) and minimising the inclusion of misclassified noise (contamination). Variable candidates displaying ambiguous probability distributions were flagged using a Quality Index ($QI$), derived from normalised Shannon entropy \citep{Shannon1948}, where a $QI$ score approaching 1.0 denotes a highly secure, unambiguous classification.

\begin{figure}[ht!]
\centering
\includegraphics[width=1\columnwidth]{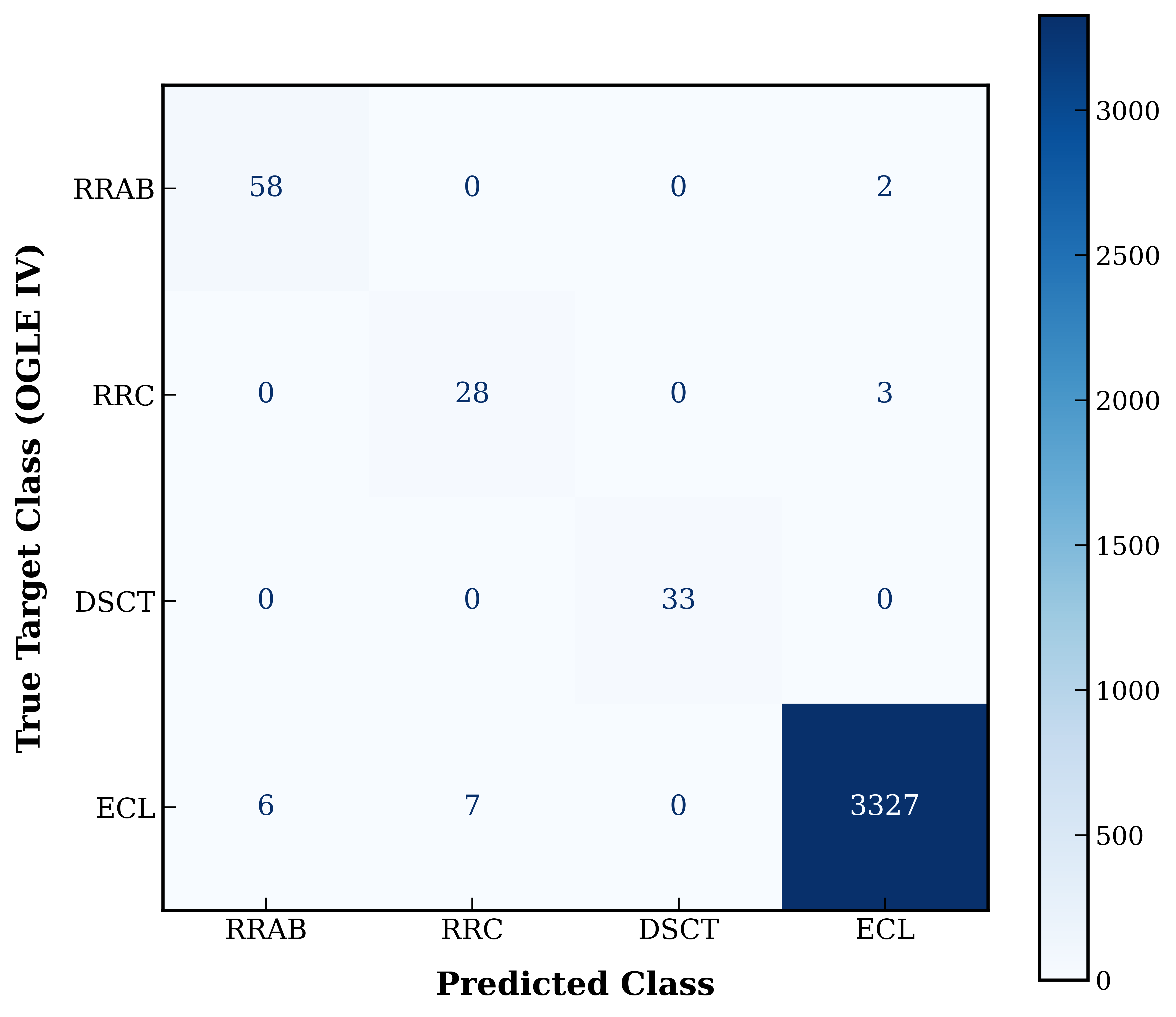}
\caption{Confusion matrix evaluated on the independent OGLE\,IV cross-matched validation sample (N = 3,464).}
\label{fig:confusion}
\end{figure}

As summarised in Table~\ref{tab:sample_distribution}, the classifier achieved an overall validation accuracy (weighted recall) of 99.48\% and a weighted precision of 99.51\%. The recovery of the DSCT and ECL classes is highly robust. The recovery of the DSCT class is 100\%, demonstrating that our short-period feature space cleanly separates these short amplitude pulsators from binary contaminants. However, it is noted that achieving 100\% on a clean 33-object (DSCT) test set does not imply the model is universally 100\% flawless on noisier, fainter, or lower amplitude external targets; This demonstrates that no classification errors occurred within this high-quality validation subset of 33 OGLE\,IV DSCT light curves. The slight reduction in RRc and RRab completeness (recall = 0.9032 and 0.9667, respectively) is a direct consequence of the strict probability thresholds designed to eliminate high-amplitude contact binary contamination. To trace individual misclassification pathways across the independent validation sample (sources in common with OGLE\,IV), the corresponding confusion matrix is presented in Fig.~\ref{fig:confusion}.

\subsection{Catalogue of variables classified using machine learning}

\begin{figure*}[ht!]
\centering
    \begin{subfigure}[t]{0.24\textwidth}
        \centering
        (a) GAGE--25\_44384\\[0.5ex] 
        \includegraphics[width=\textwidth]{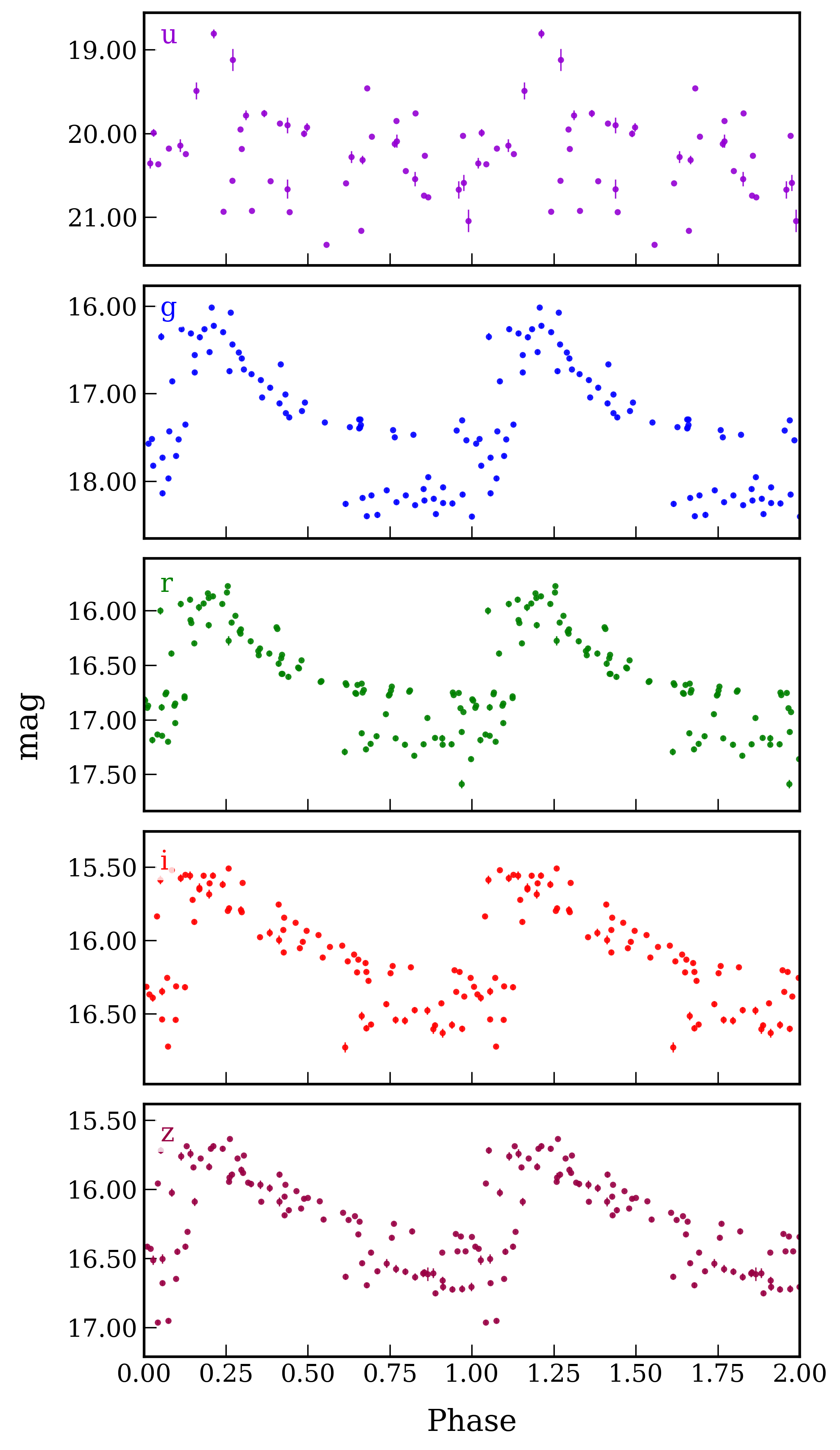}
        \label{fig:star_44384}
    \end{subfigure}
    \hfill
    \begin{subfigure}[t]{0.24\textwidth}
        \centering
        (b) GAGE--49\_415076\\[0.5ex] 
        \includegraphics[width=\textwidth]{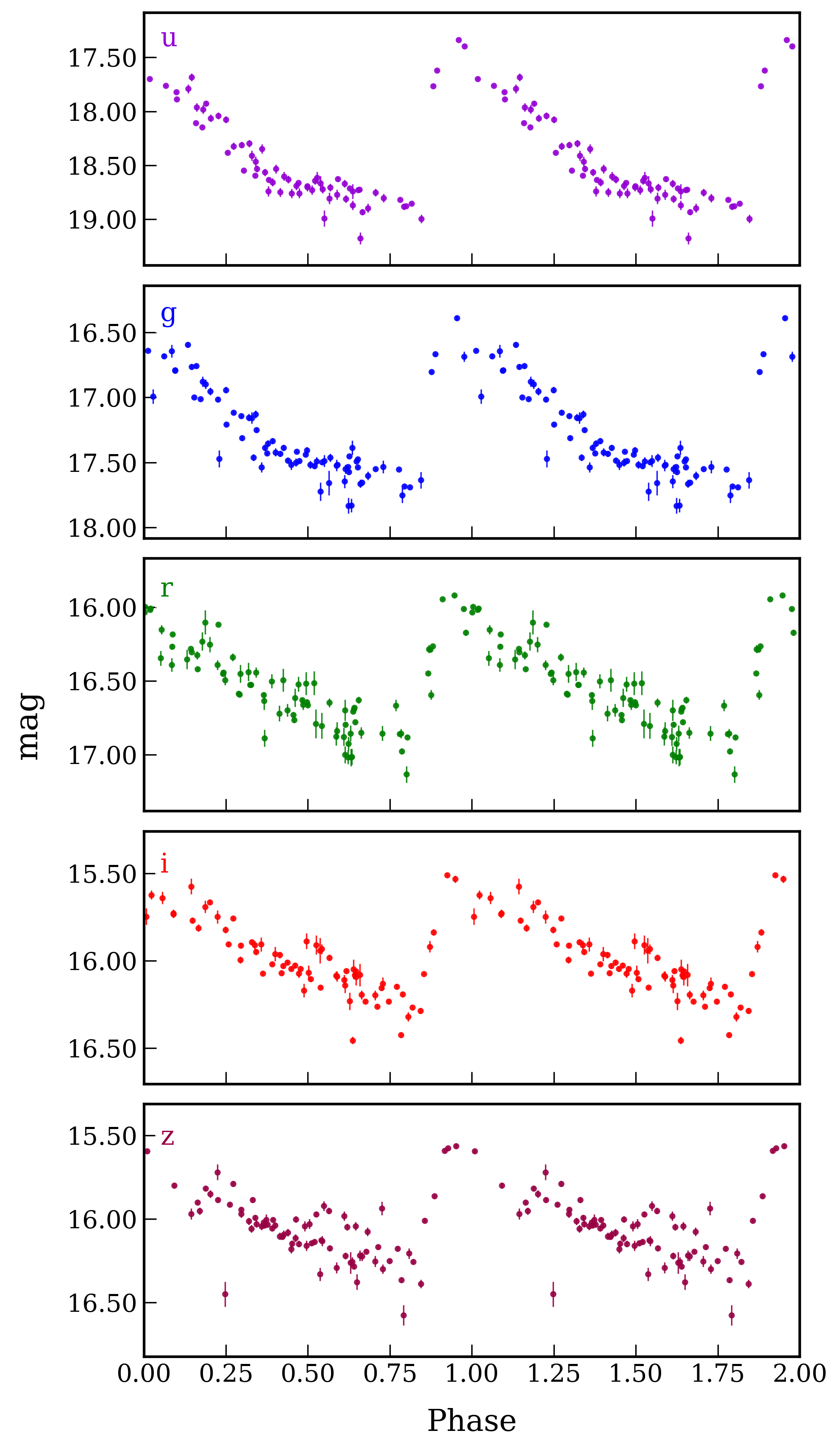}
        \label{fig:star_415076}
    \end{subfigure}
    \hfill
    \begin{subfigure}[t]{0.24\textwidth}
        \centering
        (c) GAGE--55\_204537\\[0.5ex] 
        \includegraphics[width=\textwidth]{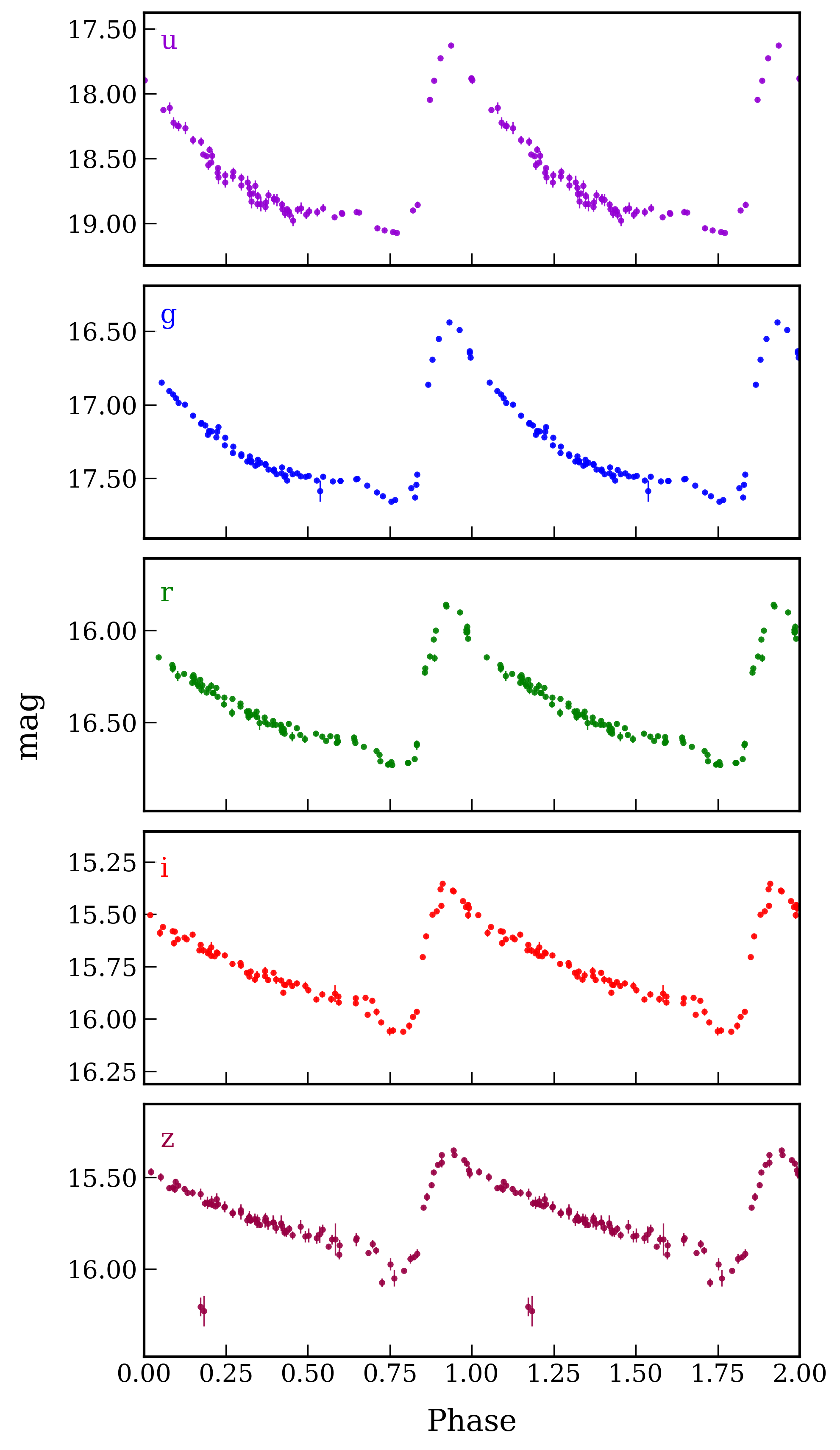}
        \label{fig:star_204537}
    \end{subfigure}
    \hfill
    \begin{subfigure}[t]{0.24\textwidth}
        \centering
        (d) GAGE--11\_349786\\[0.5ex] 
        \includegraphics[width=\textwidth]{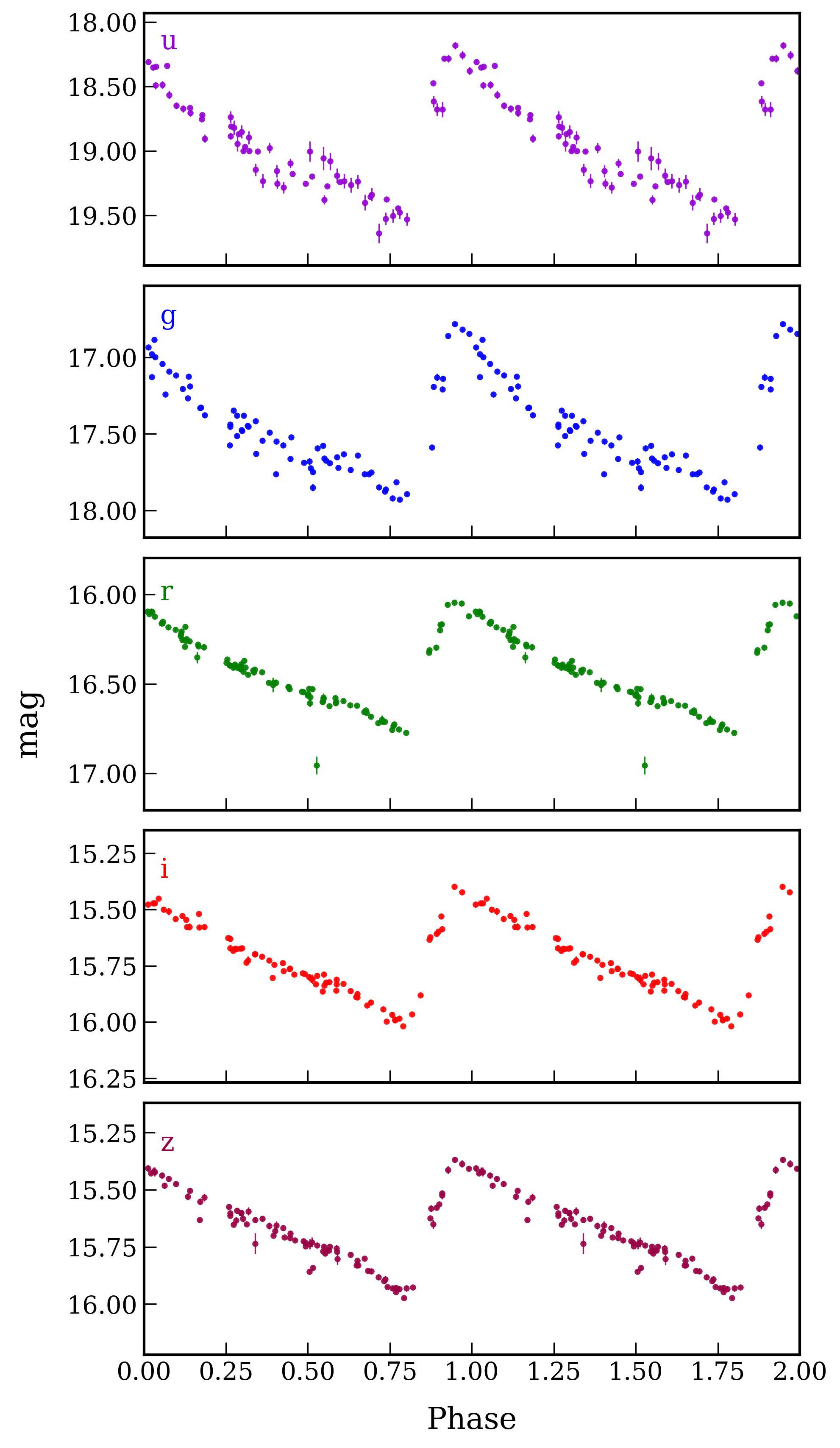}
        \label{fig:star_349786}
    \end{subfigure}
\caption{Phased multi-band ($ugriz$) DECam light curves for four representative fundamental-mode RRL (RRab) stars, sorted by discovery tier. All data points are phased to their best-fit GLS period and colour-coded by filter. Panel (a): Source GAGE--25\_44384 -- a novel variable identified and classified by this work. Panel (b): Source GAGE--49\_415076 -- a variable classified by this work and detected by VVV as well. Panel (c): Source GAGE--55\_204537 -- a variable detected by our pipeline and the \textit{Gaia}~DR3 variability catalogue but missing from OGLE\,IV. Panel (d): Source GAGE--11\_349786 -- a bright, well-sampled variable independently recovered by our work, \textit{Gaia}~DR3, and the OGLE\,IV catalogue.}
\label{fig:multiband_showcase}
\end{figure*}

Table~\ref{tab:discovery_summary}a summarises the mean physical parameters of the 1,736 newly identified variables in this work that were absent from the OGLE\,IV catalogue, with the final column listing the subset of candidates satisfying the high-confidence threshold ($QI \geq 50\%$) and Table~\ref{tab:discovery_summary}b highlighting the top ten candidates ranked by their $QI$. 
\begin{table*}[ht!]
\caption{Ensemble properties and top high-purity classified variable candidates according to our ML analysis.}
\label{tab:discovery_summary}
\centering
\begin{tabular}{lcccc}
\hline\hline
\noalign{\smallskip}
\multicolumn{5}{c}{\text{(a)} Mean physical parameters of the newly classified variable populations} \\
\noalign{\smallskip}\hline\noalign{\smallskip}
Class & Number & $\langle \text{Period} \rangle$ (d) & $\langle \text{Amp}_r \rangle$ (mag) & Number with $\text{QI} \ge 50\%$ \\ 
\noalign{\smallskip}\hline\noalign{\smallskip}
RRab & 8     & 0.5978 & 0.6743 & 5 \\
RRc  & 7     & 0.2759 & 0.2802 & 3 \\
DSCT & 101   & 0.0750 & 0.2446 & 78 \\
ECL  & 1,555 & 3.8773 & 0.2565 & 1,476 \\
\noalign{\smallskip}\hline\hline\noalign{\smallskip}
\multicolumn{5}{c}{\text{(b)} Top 10 ML classified candidates ranked by classification certainty ($\text{QI}$)} \\
\noalign{\smallskip}\hline\noalign{\smallskip}
Source ID & Period (d) & $\text{Amp}_r$ (mag) & Class & $\text{QI}$ \\ 
\noalign{\smallskip}\hline\noalign{\smallskip}
GAGE--55\_204537 & 0.529780 & 0.6877 & RRab & 0.9874 \\
GAGE--55\_121574 & 0.314954 & 0.3487 & RRc  & 0.8665 \\
GAGE--25\_199537 & 0.663676 & 0.6882 & RRab & 0.9704 \\
GAGE--45\_118833 & 0.285966 & 0.3423 & RRc  & 0.8376 \\
GAGE--49\_415076 & 0.552561 & 0.9197 & RRab & 0.9297 \\
GAGE--28\_212892 & 0.042475 & 0.1438 & DSCT & 0.8074 \\
GAGE--46\_123306 & 0.041546 & 0.1659 & DSCT & 0.9250 \\
GAGE--46\_266040 & 0.041852 & 0.1011 & DSCT & 0.6964 \\
GAGE--45\_117853 & 0.523122 & 0.5248 & RRab & 0.8842 \\
GAGE--40\_185148 & 0.064174 & 0.1171 & DSCT & 0.6851 \\
\noalign{\smallskip}\hline
\end{tabular}
\end{table*}
Using the ML classifier, we identified 15 RRL stars, including eight RRab and seven RRc with mean periods of $\langle P \rangle = 0.5978$\,d and $\langle P \rangle = 0.2759$\,d, respectively. Using ML, 1,608 candidates were initially classified as ECL. However, cross-matching with the \textit{Gaia}~DR3 catalogue and CMD analysis reclassified 53 of these sources as LPV (Sec. \ref{sec:cmd}), resulting in a finalised sample of 1,555 confirmed ECL. Additionally, we report a population of 101 DSCT candidates. These short-period pulsators have a mean period $\langle P \rangle = 0.0750$\,d. The near-unity confidence metrics of individual targets, such as the fundamental pulsator GAGE--55\_204537 ($QI=0.9874$) and the short-period candidate GAGE--46\_123306 ($QI=0.9250$), demonstrate the model's ability to cleanly isolate distinct pulsation curves near our survey boundaries. To illustrate the data quality spanning our discovery space, Fig.~\ref{fig:multiband_showcase} presents phased multi-band ($ugriz$) light curves for four representative RRab stars with $QI \geq 50\%$. 

Panel a presents a novel, low-luminosity RRab candidate classified via ML by our pipeline (GAGE--25\_44384). The phased photometry displays a classic asymmetric sawtooth profile, exhibiting a coherent reduction in peak-to-peak amplitude and a smoother morphology towards longer wavelengths ($g \rightarrow z$), as expected for pulsating stellar atmospheres \citep{Das2018}. Notably, the structured track dispersion along the rising branch and maximum light in the high signal-to-noise filters ($g, r, i, z$) strongly hints at the Blazhko effect \citep{Blazhko1907}. Conversely, the higher scatter in the $u$ band is driven by a combination of reduced detector sensitivity, atmospheric attenuation, and significantly elevated interstellar extinction at near-ultraviolet wavelengths. Compared to the brighter benchmarks in panels b, c, and d of Fig.~\ref{fig:multiband_showcase}, the recovery of this faint source directly underscores the sensitivity of our photometric pipeline down to low-luminosity limits. Figure~\ref{fig:multiband_showcase}b shows an RRab candidate (GAGE--49\_415076) classified by ML in this work. This variable is found in the VVV catalogue \citep{Molnar2022VIVACE} as RRab as well. Despite its low luminosity, the deep DECam exposures map its characteristic asymmetric sawtooth profile across all five filters, showing the pipeline's sensitivity in high-density stellar regions. Panel c tracks a target (GAGE--55\_204537) detected by our pipeline and the \textit{Gaia}~DR3 catalogue but not found in the OGLE\,IV catalogue. Panel d displays a bright reference variable (GAGE--11\_349786) captured by all three surveys. The comprehensive multi-band photometric catalogue comprising all 6,304 high-quality variable candidates is made publicly available at the Centre de Données Astronomiques de Strasbourg (CDS); the full data schema and column descriptions are detailed in Table~\ref{tab:variables}. For the subset of 1,736 newly classified variables lacking OGLE\,IV matches, external cross-matches with \textit{Gaia}~DR3 and VVV were integrated into the catalogue. Moreover, for LPV class, the final classification (\texttt{class\_GAGE}) incorporates visual and CMD inspections. Objects classified as LPV in \textit{Gaia}~DR3 are adopted as \texttt{LPV} with \texttt{class\_source} = \texttt{Gaia\_LPV}. Uncatalogued red candidates with tentative periods are designated as \texttt{LPV\_candidate} with \texttt{class\_source} = \texttt{CMD\_LPV} to facilitate follow-up studies. Candidates with insufficient photometric or temporal constraints to yield reliable ML classification are labelled as \texttt{UNCLASSIFIED}.

\section{Results and discussion}
\label{sec:results}
\subsection{Colour--magnitude diagrams} \label{sec:cmd}

\begin{figure*}[ht!]
  \sidecaption
  \includegraphics[width=5.8cm]{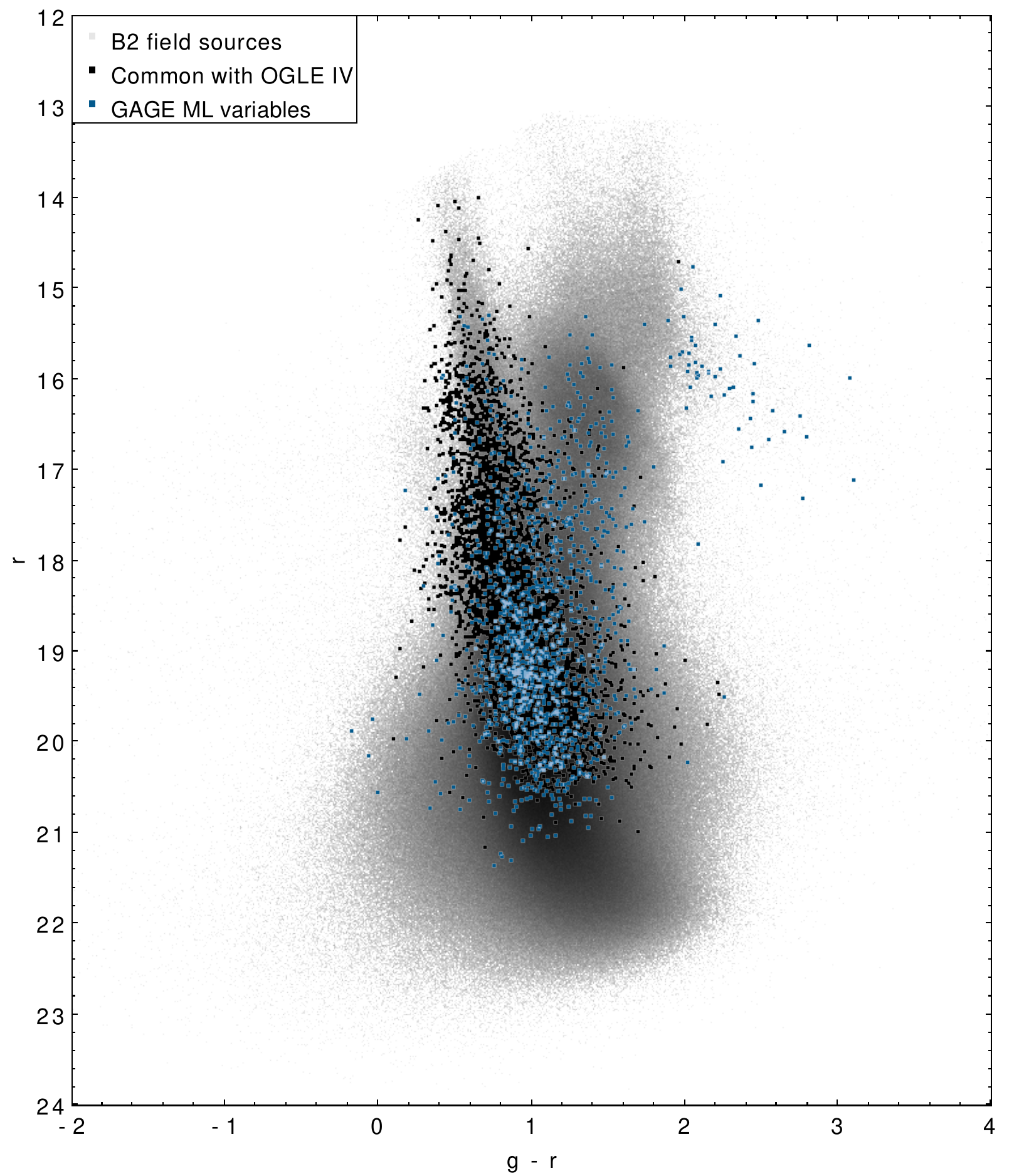}\hfill
  \includegraphics[width=5.8cm]{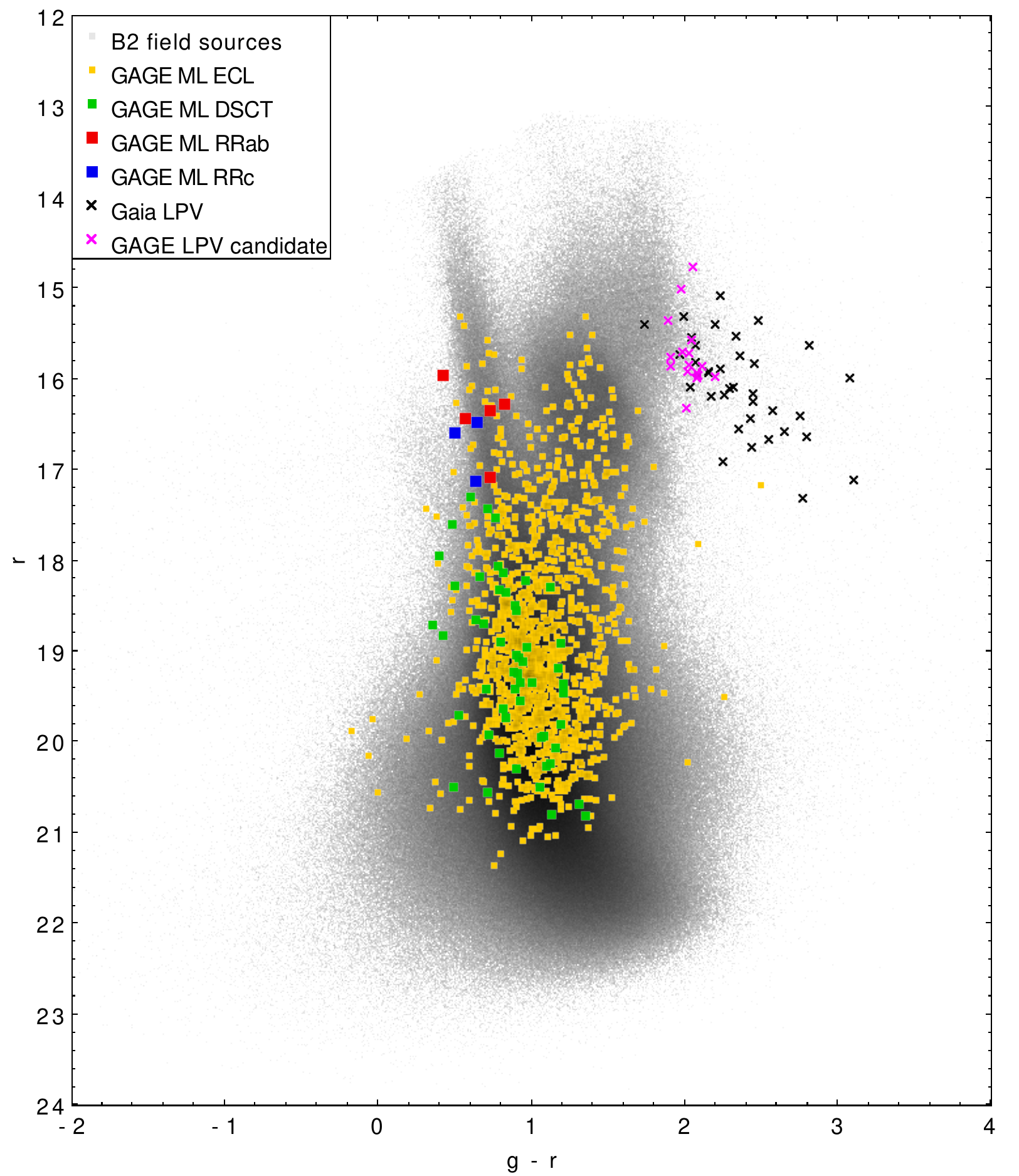}
  \caption{Colour-magnitude diagrams in the optical $SDSS$ bands ($r$ vs $g-r$). The grey density map represents the stellar population of the MW B2 field. \text{Panel (a):} Spatial distribution of variable candidates classified via the ML in this work (blue density squares) alongside variables recovered in the OGLE\,IV catalogue (black squares), plotted regardless of their quality index ($QI$). \text{Panel (b):} Specific classification distribution of the high-confidence sub-sample ($QI \geq 50\%$). Newly classified fundamental-mode RRLs (RRab; red squares), first-overtone RRLs (RRc; blue squares), $\delta$~Scuti stars (DSCT; green squares), and ECLs (ECL; yellow squares) are explicitly highlighted. Additionally, Gaia\_LPV and LPV candidates are marked with black and magenta crosses, respectively.}
  \label{fig:cmd_main_optical}
\end{figure*}

Figure~\ref{fig:cmd_main_optical} displays the optical $r$ versus $(g-r)$ colour--magnitude diagram (CMD) of the targeted Baade's window B2 field. Our newly classified variable candidates via ML (not previously detected by OGLE) and those in common with the OGLE\,IV catalogue are over-plotted against the stellar population of our survey footprint.

As shown in Fig.~\ref{fig:cmd_main_optical}a, the variable sources closely trace the primary evolutionary sequences of the stellar field. The cross-matched OGLE\,IV variables (black points) trace the empirical boundaries of the RRL instability strip as well as the luminous ECL sequences, with the latter following the main sequence and the brighter blue plume on the CMD. Additionally, the newly identified variable candidates (blue density features) extend significantly deeper down to $r \sim 21.5$\,mag, rather than those in common with OGLE\,IV. In particular, it can be seen in Fig.~\ref{fig:cmd_main_optical}b that the positions of the newly classified variable stars in the CMD are in agreement with their classification.

Figure~\ref{fig:cmd_main_optical}b maps the distribution of the high-confidence subset ($QI \geq 50\%$) that is absent from the OGLE\,IV catalogue (comprising 1,736 out of 6,304 variable candidates detected in this work), classified via the ML infrastructure detailed in Sect.~\ref{sec:ml_methods}. This un-catalogued population spans a broad multi-band range, tracking from faint ultraviolet thresholds ($u\approx 17.4\text{--}22.6$\,mag) to deeper near-infrared limits ($z\approx 14.7\text{--}20.7$\,mag).

\begin{figure*}[ht!]
  \sidecaption
  \includegraphics[width=5.8cm]{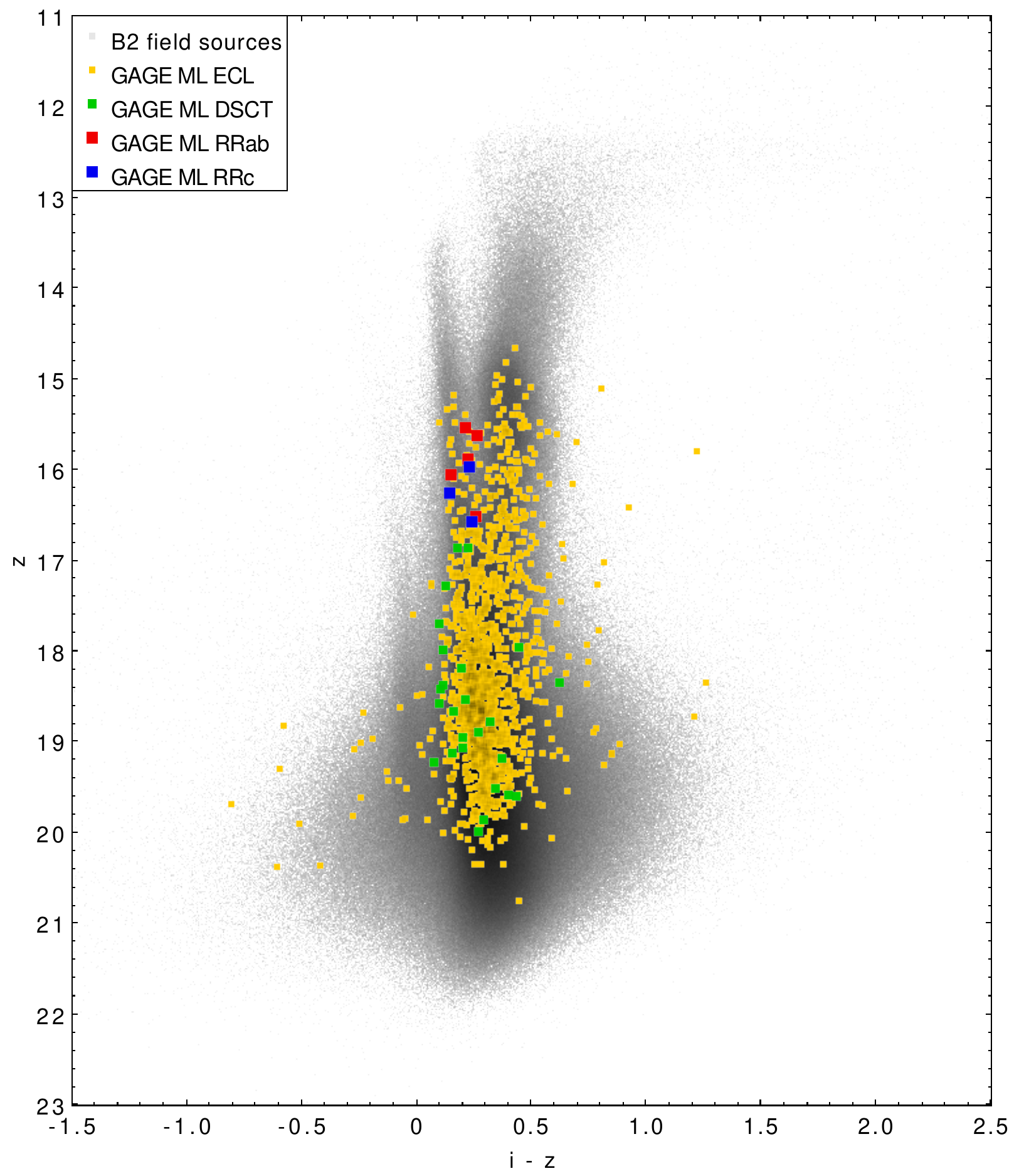}\hfill
  \includegraphics[width=5.8cm]{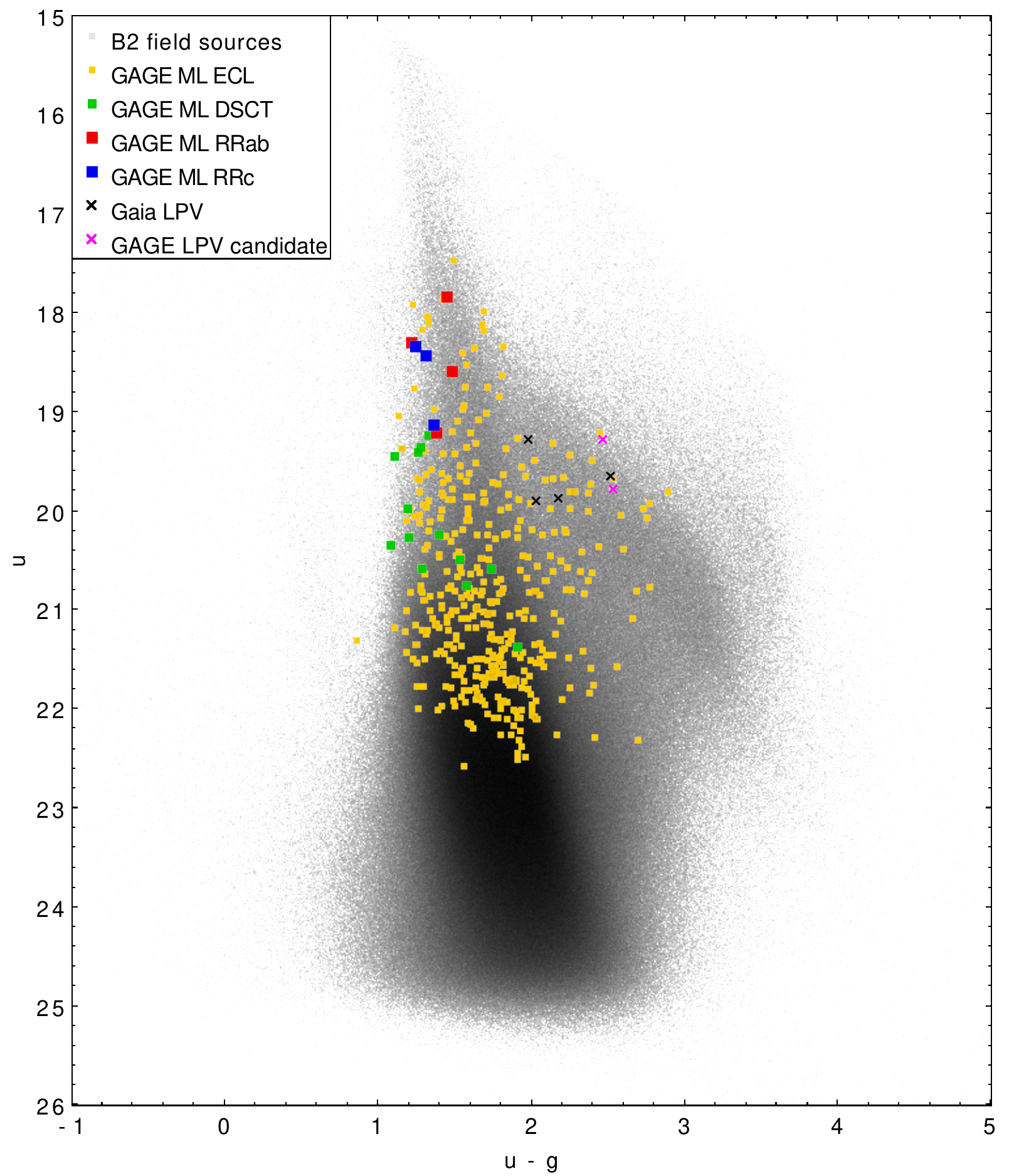}
  \caption{Multi-band CMDs for all newly classified variable candidates with $QI \ge 50\%$ (colour-coded) plotted against the B2 field stellar population (grey density). \text{Panel (a):} Near-infrared $z$ vs $(i-z)$ diagram in which the tight vertical alignment of the variables reflects the reduced impact of interstellar extinction at longer wavelengths. \text{Panel (b):} Near-ultraviolet $u$ vs $(u-g)$ diagram. The variables trace the hot ultraviolet-bright stellar sequences, showcasing the depth of DECam's $u$ band in highly dust-attenuated Bulge fields. The majority of the confirmed LPV (black crosses) and LPV candidates (magenta crosses) are omitted from these panels due to a lack of valid photometric measurements in the specific bands plotted.}
  \label{fig:cmd_infrared_uv}
\end{figure*}

An \text{a-posteriori} cross-matching of our 6,304 variable candidates against external catalogues reveals 1,984 sources in common with \textit{Gaia}~DR3 \citep{Rimoldini2023_Variability} and 2,496 with VVV \citep{Molnar2022VIVACE}. Restricting this comparison to our 1,736 newly classified variable candidates, we found 198 sources in common with \textit{Gaia}~DR3 and 230 with VVV. Within the \textit{Gaia} overlap, 159 sources ($80.30\%$) share identical classification labels (ECL and RRL), with 143 of these matching targets meeting our QI threshold ($QI \geq 50\%$). For the VVV subset, 218 sources ($94.8\%$) show classification agreement across all three ML classes (ECL, DSCT, RRL), 198 of which have $QI \geq 50\%$. Overall, our ML model initially classified 1,585 of the 1,736 newly identified variables with $QI \geq 50\%$, including 166 sources overlapping with \textit{Gaia} (83.8\%) and 208 overlapping with VVV (90.4\%). This count of 1,585 high-confidence candidates was subsequently adjusted to 1,562 after 23 sources classified as ECL were re-identified as LPV based on \textit{Gaia} classifications and CMD analysis, as detailed at the end of this section. Finally, within our sample of 1,736 newly classified variables, our pipeline identifies a robust subset of 1,391 previously unknown candidates, 1,282 of which meet our high-confidence threshold ($QI \geq 50\%$). These sources are absent from the OGLE\,IV, VVV, and \textit{Gaia}~DR3 catalogues and are predominantly concentrated in the faint regime of the CMD. This explicit stratification confirms that our new discoveries are primarily a consequence of our deeper photometric baseline, rather than positional or observational omissions by previous wide-field surveys such as OGLE\,IV, \textit{Gaia}~DR3, or VVV. The 1,282 high-confidence sample of new variable candidates classified via ML comprises 1,203 ECLs, 77 DSCTs, and two RRc stars. A minor fraction of the newly classified sample (11 out of 1,736 candidates) could not be reliably classified by our ML model due to insufficient observational data. Furthermore, it should be noted that due to the severe numerical scarcity of certain variable types within the OGLE\,IV sources in common with our variable list, specific macro classes could not be incorporated into the training phase. This structural limitation impacted two distinct populations:
\begin{enumerate}
    \item \text{Cepheids}: Only six baseline sources were available in the training set, causing the ensemble classifier to systematically misassign unmodelled Cepheid candidates to the morphologically similar ECL class.
    \item \text{LPVs}: No baseline LPV sources were present within our targeted footprint, leaving the classifier without LPV training examples. Consequently, true LPV candidates were initially absorbed into the ECL population.
\end{enumerate}
To recover these suppressed populations, we performed a targeted inspection of 54 candidates initially classified as ECL that exhibit red colours (g--r) exceeding $\approx 2$\,mag on the CMD (Fig.~\ref{fig:cmd_main_optical}b). Among these, the \textit{Gaia}~DR3 variability catalogue classified 37 objects as LPV stars; these sources are accordingly assigned a final classification of \texttt{LPV} in our catalogue (black crosses in Fig.~\ref{fig:cmd_main_optical}b) . For the remaining 17 targets unclassified by \textit{Gaia}, we designated 16 of them as LPV candidates (magenta crosses) based on their distinct location on the CMD (Table~\ref{tab:variables}), and one was classified as binary in VVV catalogue, so we finalised its classification as ECL. The new classified RRab pulsators (red filled squares) and RRc sources (blue filled squares) cluster tightly within the expected boundaries of the instability strip between $16 < r \lesssim 17$\,mag and $0.5 \lesssim (g-r) \lesssim 1.0$\,mag. The DSCT candidates (green squares) mostly occupy the fainter main-sequence turn-off (MSTO) region at $r > 19$\,mag, whereas the newly classified ECLs (yellow squares) are distributed broadly across both the main sequence, red-giant branch, and red clump, mirroring the structural stellar density profile of the inner disk and bulge. To fully exploit the multi-band baseline of our survey, Fig.~\ref{fig:cmd_infrared_uv} presents the CMDs extended into the near-infrared ($z$ versus $i-z$) and near-ultraviolet ($u$ versus $u-g$) regimes for all newly classified variables with $QI \geq 50\%$.

The near-infrared CMD ($z$ versus $i-z$; Fig.~\ref{fig:cmd_infrared_uv}a) exhibits a remarkably sharp and well-defined stellar locus. Because longer wavelengths are significantly less affected by interstellar dust extinction, the variables line up in a distinct vertical structure. Conversely, the near-ultraviolet CMD ($u$ versus $u-g$; Fig.~\ref{fig:cmd_infrared_uv}b) shows the high-energy properties of our sample. Despite the severe dust attenuation typically blocking UV light towards the inner Galaxy, the 4--m Blanco telescope's collecting area allows the tracking of these variable candidates down to $u \sim 23.5$~mag. The broader colour dispersion in $(u-g)$ reflects the high sensitivity of the ultraviolet band to metallicities and temperature variations, providing an essential leverage for future chemical characterisation of these pulsators.

\subsection{Period distribution} \label{subsec:period_dist}

\begin{figure}[ht!]
    \centering
    \includegraphics[width=\columnwidth]{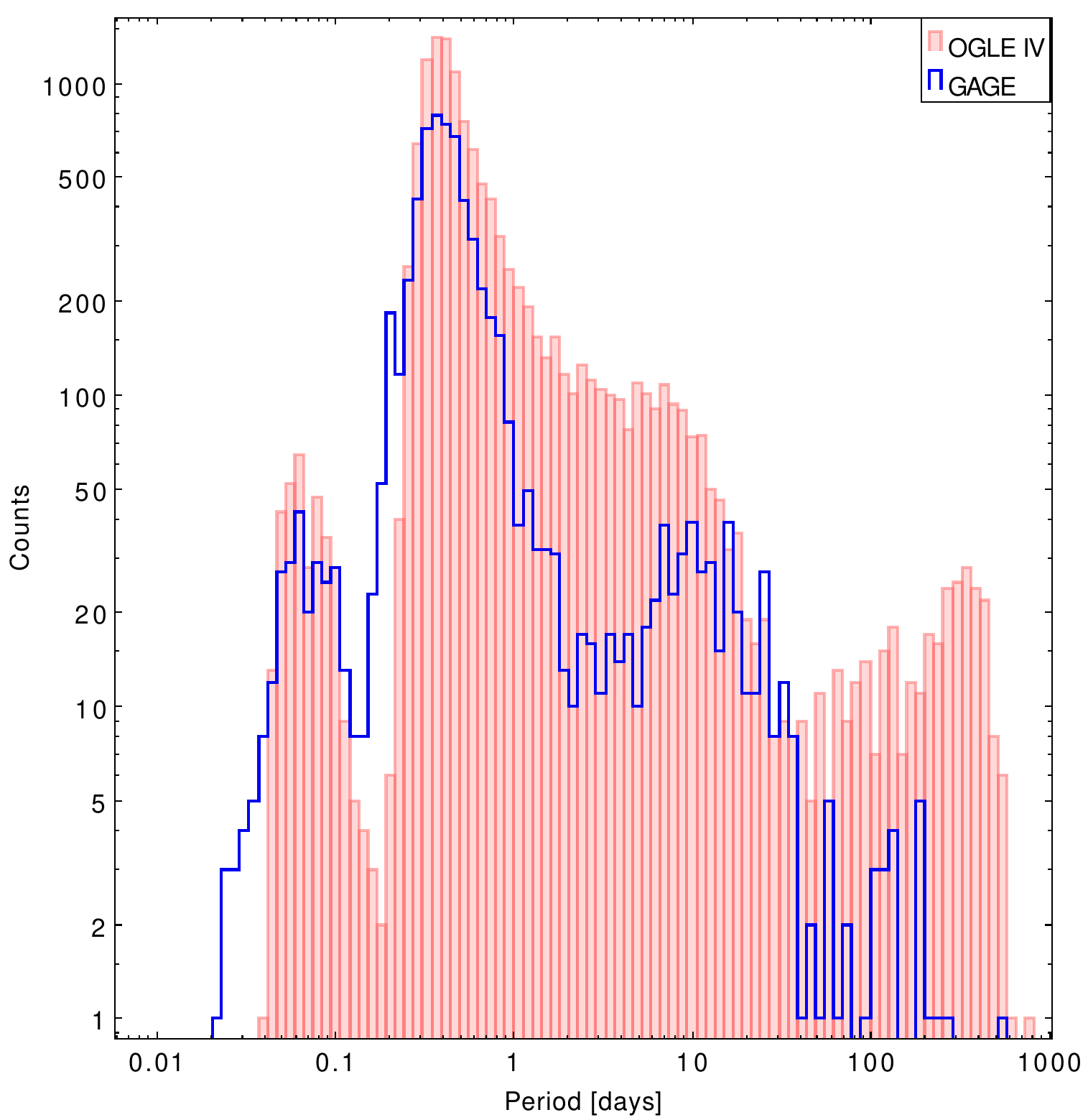}
    \caption{Distribution of calculated periods for the $N=6,304$ high-quality GAGE variable stars (blue open histogram) compared to known OGLE\,IV variables within the same footprint (pink filled bars). The prominent peak at 0.2--1.0\,d represents the core Galactic bulge RRL population, while the excess at periods $\lesssim 0.15$\,d highlights the depth of our DECam pipeline in capturing faint DSCTs and contact binaries. Conversely, the deficit between 2 and 30\,d and the sharp truncation at $>30$\,d are driven by seasonal and day-night gaps in the archival data, which create aliases and phase gaps that are rejected by our strict RANSAC quality filters.}
    \label{fig:period}
\end{figure}

Figure~\ref{fig:period} presents the period histogram of the GAGE detected variables in this work (blue open histogram) in direct comparison with the recovered OGLE\,IV variables within the same field footprint (pink filled bars). The period axis is presented on a logarithmic scale with linear day units to effectively map the wide range of stellar periodicities. The prominent, well-documented peak centred between $0.2$\,d and $1.0$\,d closely matches the core Galactic bulge RRL population, validating both the photometric precision and the period-finding accuracy of our GLS pipeline. Crucially, our deep DECam photometry reveals a significant statistical excess of very short period variables extending down to $\sim 0.02$\,d. This prominent feature highlights our survey's ability to probe low-amplitude DSCT populations and faint contact binaries down to $r \sim 21.5$\,mag, effectively extending the completeness limits of shallower wide-field surveys in highly crowded fields.
Conversely, a sharp truncation occurs at periods above 30 days, where OGLE\,IV contains numerous LPVs. We also recover fewer variables between 2 and 30 days. This intermediate deficit is driven by seasonal gaps and day-night cycles in the ground-based DECam archival data, which create strong aliases and disrupt phase coherence. Because our strict RANSAC criteria requires uniform phase coverage to guarantee catalogue purity, both these poorly sampled intermediate sources and LPVs are systematically filtered out.

\subsection{Magnitude distribution}\label{subsec:mag}

\begin{figure}[ht!]
    \centering
    \includegraphics[width=\columnwidth]{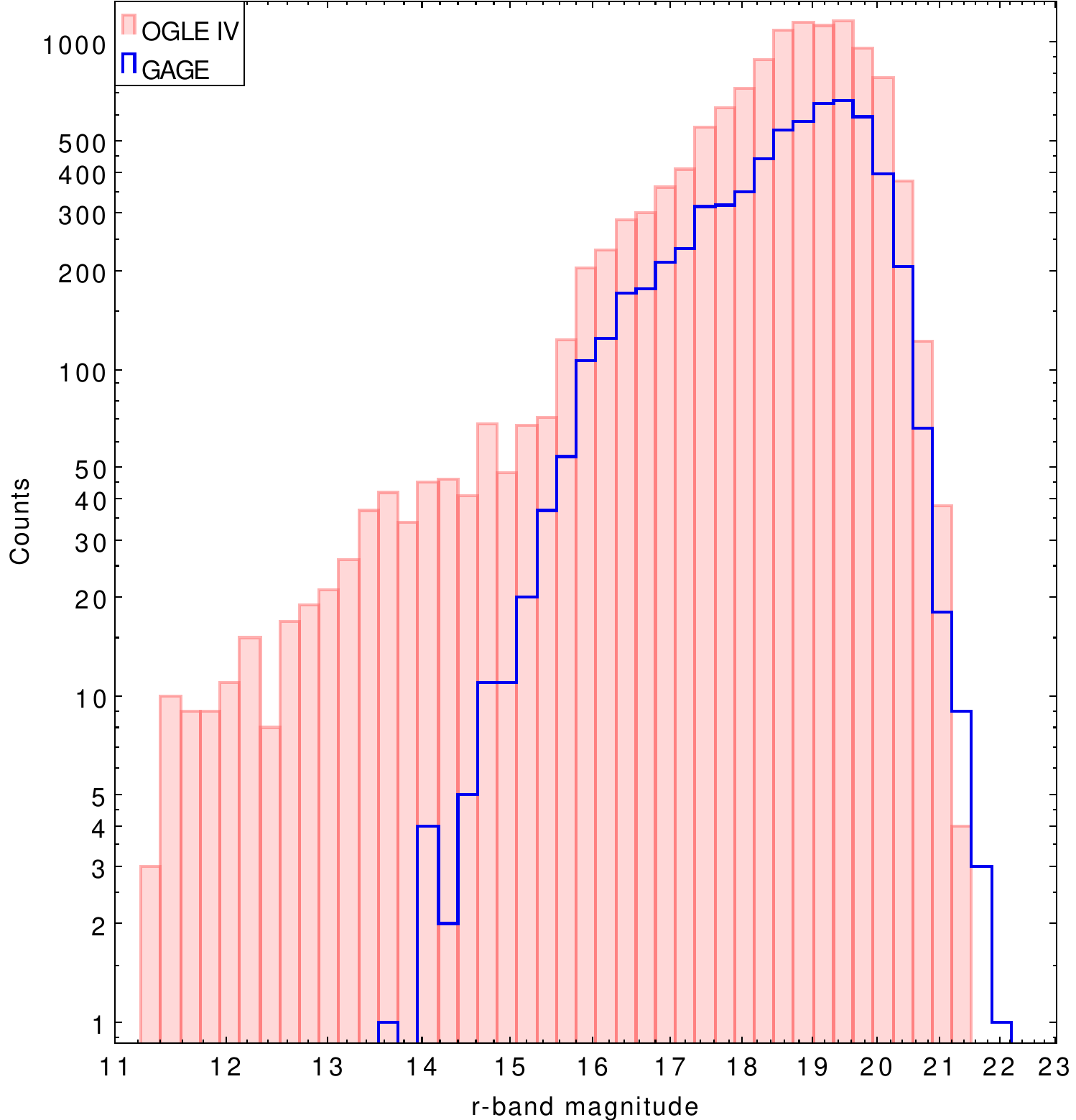}
    \caption{$r$--band magnitude distribution of the $N=6,304$ HQ GAGE variables (solid~blue line) compared with the transformed OGLE\,IV sample in B2 (pink filled bars). OGLE\,IV $I$--band magnitudes were mapped into the DECam filter space via the empirical linear relation derived in Eq.~\ref{eq:phot_transform}. The excellent agreement across the $16 < r < 20.5$\,mag regime validates our cross-filter calibration. The GAGE census demonstrates clear completeness advantages approaching the faint detection limit ($r \sim 22$\,mag), while its bright-end truncation ($r < 14$\,mag) marks the saturation boundary of our deep public DECam exposures.}
    \label{fig:mag_dist}
\end{figure}

To compare the photometric depth of our detected sample directly with the full OGLE\,IV baseline (including the OGLE\,IV variables not detected by our pipeline but located within our footprint), we mapped the OGLE\,IV $I$--band magnitudes into the $SDSS$ $r$--band filter space. Using a representative subset of HQ cross-matched stars distributed across our footprint, we derived an empirical linear transformation parametrising the colour offset ($r - I_{\mathrm{OGLE}}$) as a function of the $r$--band magnitude. An ordinary least-squares regression yields the following conversion relation:
\begin{equation}
    r - I_{\rm OGLE} = 0.1331 \times r - 1.5419
    \label{eq:phot_transform},
\end{equation}
where $r$ represents our calibrated $SDSS$ $r$--band magnitude and $I_{\mathrm{OGLE}}$ denotes the corresponding OGLE\,IV $I$--band magnitude. The linear fit yields a correlation coefficient of $R = 0.69$ with an RMS deviation of $\sigma = 0.096$~mag. This relation was subsequently applied to homogenise the baseline catalogue fields. Figure~\ref{fig:mag_dist} shows the $r$--band magnitude distribution for both the GAGE variables (blue open histogram) and the transformed OGLE\,IV catalogue within B2 field (pink filled bars). The distributions match closely across the peak stellar density regime between 16 and 20.5\,mag, confirming the precision of our cross-filter calibration and demonstrating that our catalogue achieves a deeper photometric limit than the OGLE\,IV survey.
At the bright end ($r < 14$\,mag), the GAGE distribution shows a sharp drop-off primarily driven by a strict saturation threshold enforced during our image-level quality control. Conversely, at the faint limit, the GAGE census extends down to $r \sim 22$\,mag, smoothly tracing populations roughly 0.5 to 1.0 magnitude deeper than the limiting detection threshold where the transformed OGLE\,IV completeness rapidly decays.

\subsection{Amplitude distribution}
\label{subsec:amplitude}
\begin{figure}[ht!]
\centering
\includegraphics[width=\columnwidth]{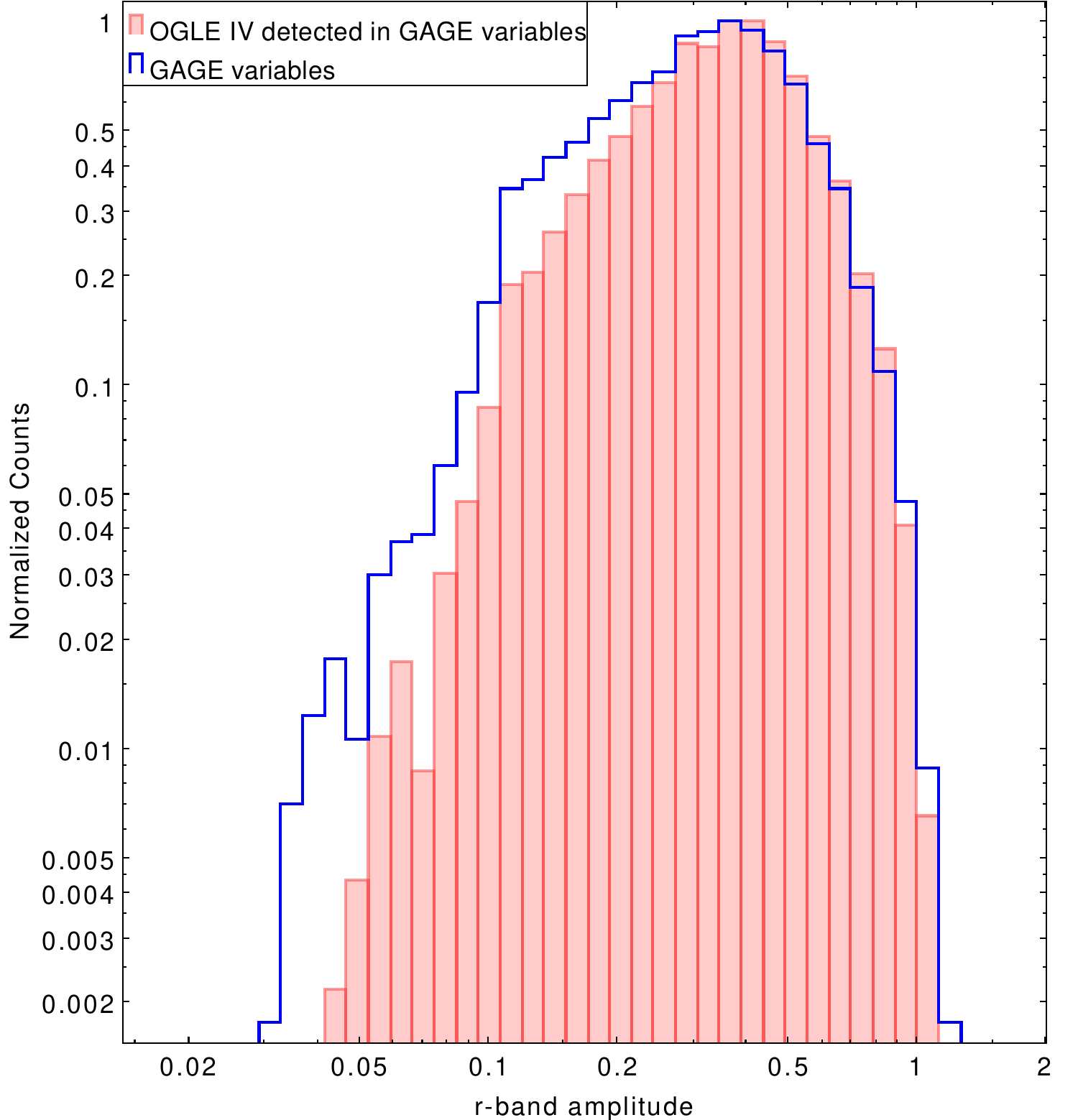}
\caption{Normalised histogram of RANSAC-derived $r$--band amplitudes for our HQ variable star sample (blue open histogram) compared with the subset also detected by OGLE\,IV (pink filled bars). In both distributions, the primary peak at $A_r \approx 0.35$\,mag is driven by the central population of bulge RRL variables. The smooth extended tail down below $0.15$\,mag validates the sensitivity of our pipeline in recovering low-amplitude ECL and DSCT stars despite severe crowding constraints. The excess population in the GAGE sample relative to the OGLE\,IV distribution at these lower limits highlights our method's capacity to discover subtle short-amplitude variable candidates, which is likely due to enhanced amplitude detection thresholds in our native $r$ band compared to OGLE's primary $I$--band framework.}
\label{fig:amplitude_dist}
\end{figure}

The distribution of the $r$--band variability amplitudes derived from our best-fit RANSAC models is presented in Fig.~\ref{fig:amplitude_dist}. The sample exhibits a continuous distribution spanning from our effective short amplitude detection limit at $\sim 0.03$\,mag up to large amplitude pulsators exceeding $1.0$\,mag. The overall morphology is characterised by a single prominent, asymmetric peak centred at $A_r \approx 0.35$\,mag, aligning precisely with the expected maximum for the dominant fundamental-mode RRL population in the Galactic bulge. Below $A_r \sim 0.15$\,mag, the distribution forms an extended, shallow short amplitude tail. This regime captures a mixture of contact binaries, detached eclipsing systems, and short period periodic pulsators such as DSCT stars. Additionally, Fig.~\ref{fig:amplitude_dist} reveals an excess population of short amplitude variable candidates identified by GAGE that are missing from the OGLE\,IV. This divergence stems from a fundamental wavelength effect: while OGLE\,IV relies on the redder $I$--band as its primary baseline, GAGE targets these fields in the $r$ band. At these shorter wavelengths, the intrinsic flux variations of short luminosity variables (e.g. ECL and DSCT) are physically larger and more pronounced. Consequently, our $r$--band baseline lowers the effective detection threshold, unlocking a population of subtle, short amplitude variables that escape detection in longer wavelength surveys.

\subsection{The period--amplitude distribution of RR Lyrae and $\delta$ Scuti stars} \label{subsec:pa_depth}
\begin{figure}[ht!]
    \centering
    \includegraphics[width=\columnwidth]{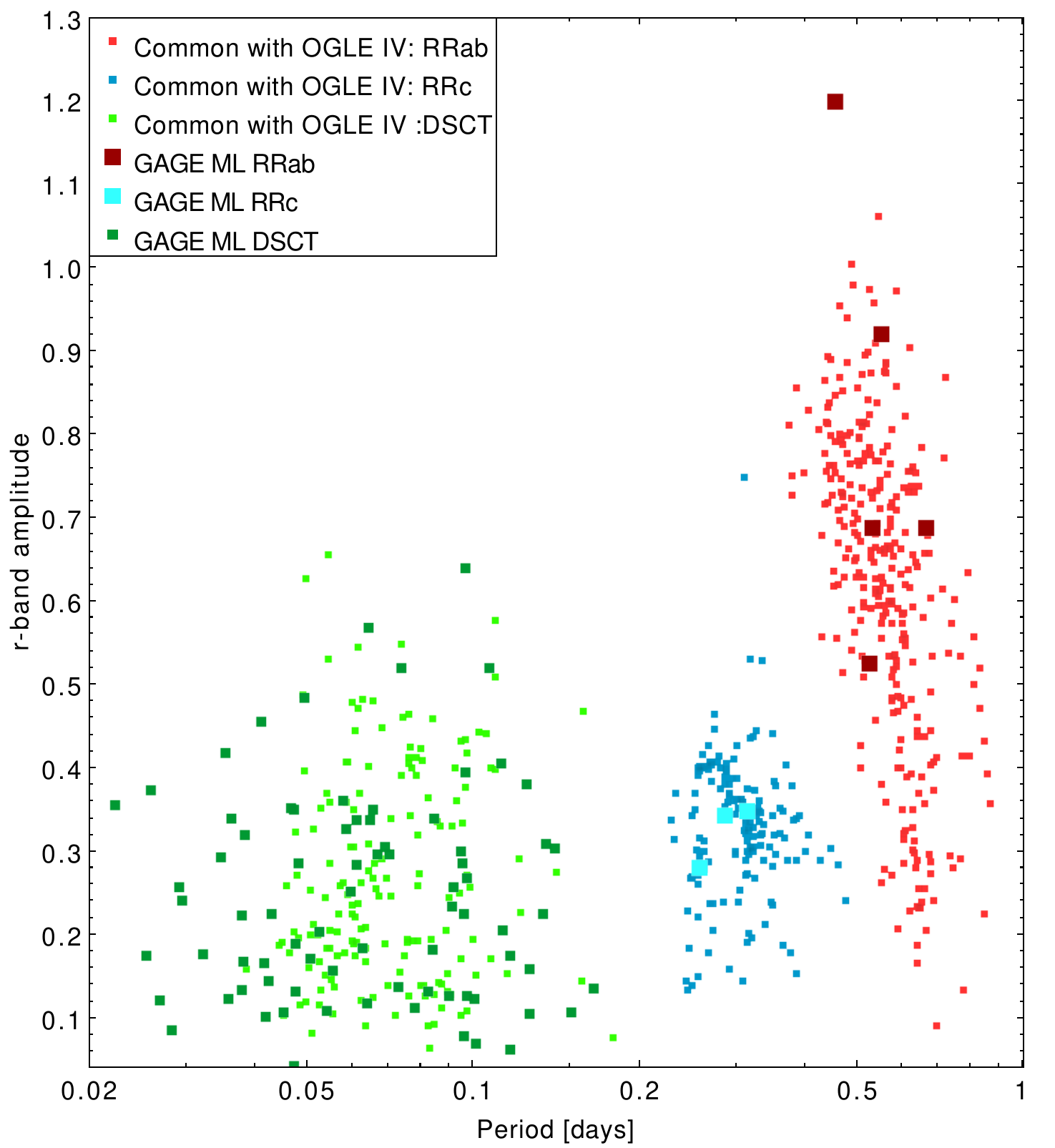}
    \caption{$r$--band period--amplitude diagram for the detected pulsating variable stars in B2. Previously catalogued variables in OGLE\,IV in common with our sample are shown in light green (DSCT), light red (RRab), and light blue (RRc). Newly ML classified variables with $QI \geq 50\%$ are explicitly highlighted as dark red squares (ML RRab), cyan squares (ML RRc), and dark green squares (ML DSCT). The clear structural separation between the distinct pulsation modes underscores the high fidelity of our ML classification framework.}
    \label{fig:bailey}
\end{figure}

To evaluate the consistency of our ML classifications, we map the high-quality short-period pulsating stars ($QI \geq 50\%$) onto the $r$--band period--amplitude diagram (Fig.~\ref{fig:bailey}). The distinct physical clustering of these variability classes across both diagnostic planes provides strong empirical validation of our method, demonstrating negligible cross-class contamination while tracing the physical structures expected in the inner Galaxy \citep[see e.g.][]{Minniti2016,saha2019}.
The RRab variables form a sharp, heavily populated locus between $0.5 \lesssim P \lesssim 0.9$\,d, tracking the classic asymmetric Bailey sequence where pulsation amplitudes systematically increase towards shorter period thresholds \citep{Bono1997, Braga2015}. Our ML classified RRab sources (dark red squares) tightly follow this sequence, including a large-amplitude candidate reaching $A_r \approx 1.2$\,mag that lacks a counterpart in both the OGLE\,IV and \textit{Gaia}~DR3 catalogues. Separated from the fundamental-mode pulsators, the first-overtone (RRc) stars predominantly occupy a compact clump between $0.23 \lesssim P \lesssim 0.45$\,d. Their lower amplitudes are constrained between 0.15 and 0.5\,mag, mapping along the characteristic bell-shaped pattern of RRc populations in period--amplitude diagrams \citep[see][]{Bono1997, Marconi2021, Marconi2022}.

We note three cross-matched RRc candidates displaying anomalously high amplitudes relative to the core overtone sequence, which may indicate blended profiles or unrecognised multi-periodic behaviour \citep{Netzel2019}. The newly ML classified RRc stars (cyan squares) smoothly populate the core of this mode.

At short periods ($P < 0.2$\,d), the distribution shows the DSCT candidates, extending down to our detection floor near $A_r \sim 0.03$\,mag (Fig.~\ref{fig:bailey}). Initial ML classification exhibited a prominent vertical feature centred at $P \sim 0.15$\,d. \cite{Chen2020_ApJS} showed that such loci around $P \sim 0.15$\,d near $r \sim 19.0$--$21.5$\,mag ($i \sim 18.5$--$20.5$\,mag) are typically contaminated by short-period contact binaries (EW) recovered at half their orbital period ($P_{\mathrm{orb}} \sim 0.30$\,d) due to surface asymmetries and GLS harmonics.

To disentangle these populations, we conducted a multi-diagnostic analysis following \cite{Chen2020_ApJS}: evaluating period--absolute Wesenheit positions \citep{Chen2020_ApJS}, Fourier phase combinations, and $g$-to-$r$ amplitude ratios ($A_g/A_r$), alongside visual phased light-curve inspection. Then, for overlapping sources in \textit{Gaia}~DR3 and VVV, period aliasing was resolved by adopting their true catalogue orbital periods. Because GLS periodograms frequently detect half-period harmonics for binary systems, short-period ECLs are particularly prone to being misclassified as DSCT pulsators, so, we re-evaluated their periods by comparing our GLS estimates directly against the periods published in the \textit{Gaia}~DR3 Variability table and the VVV multi-epoch catalogues, adopting the true orbital period whenever a half-period harmonic was identified. Re-running our ML classification with these corrected periods and diagnostic analysis reclassified the majority of the $P \sim 0.15$\,d sources as ECL.

\begin{figure*}[ht!]
  \sidecaption
  \includegraphics[width=5.8cm]{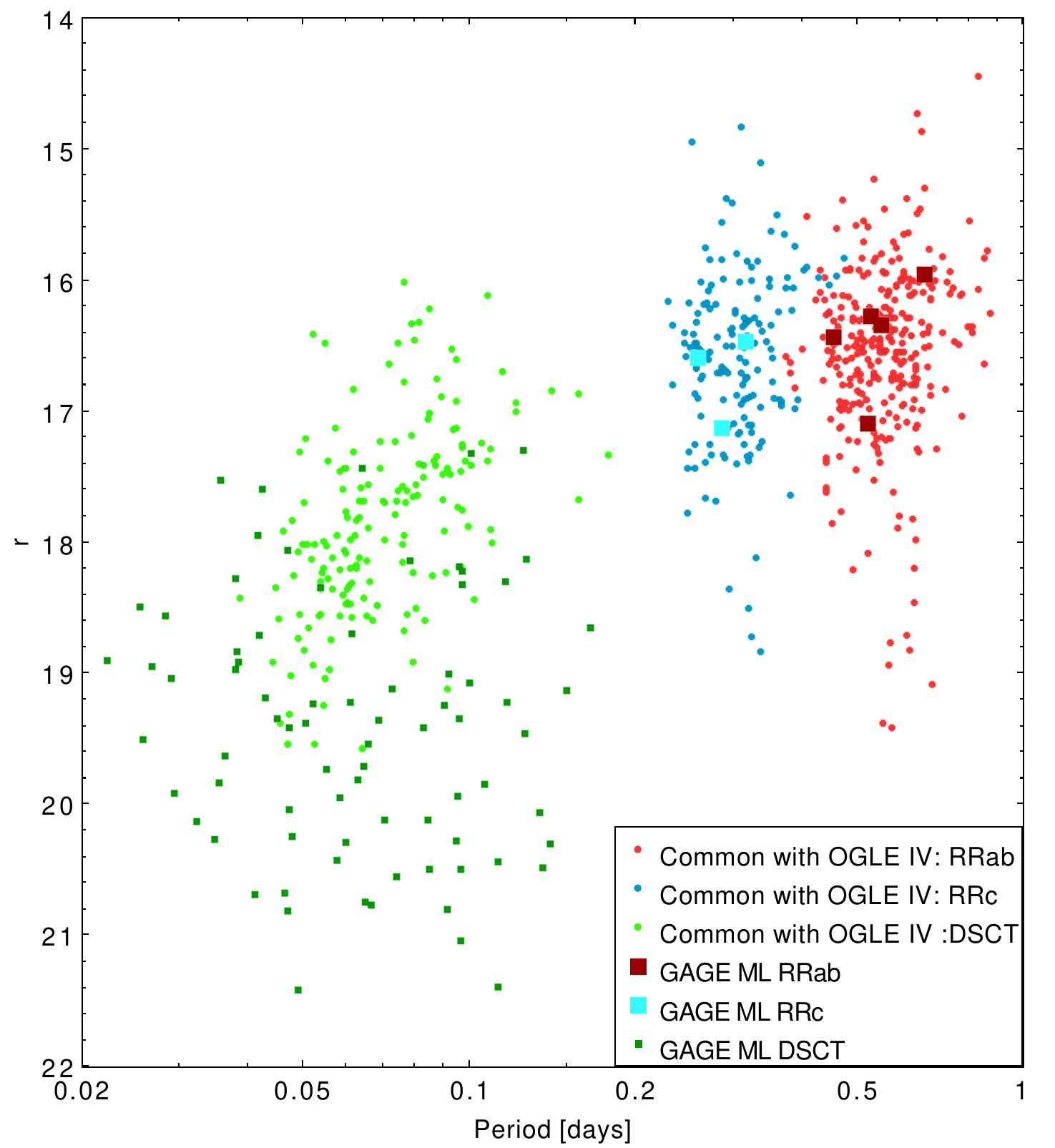}\hfill
  \includegraphics[width=5.8cm]{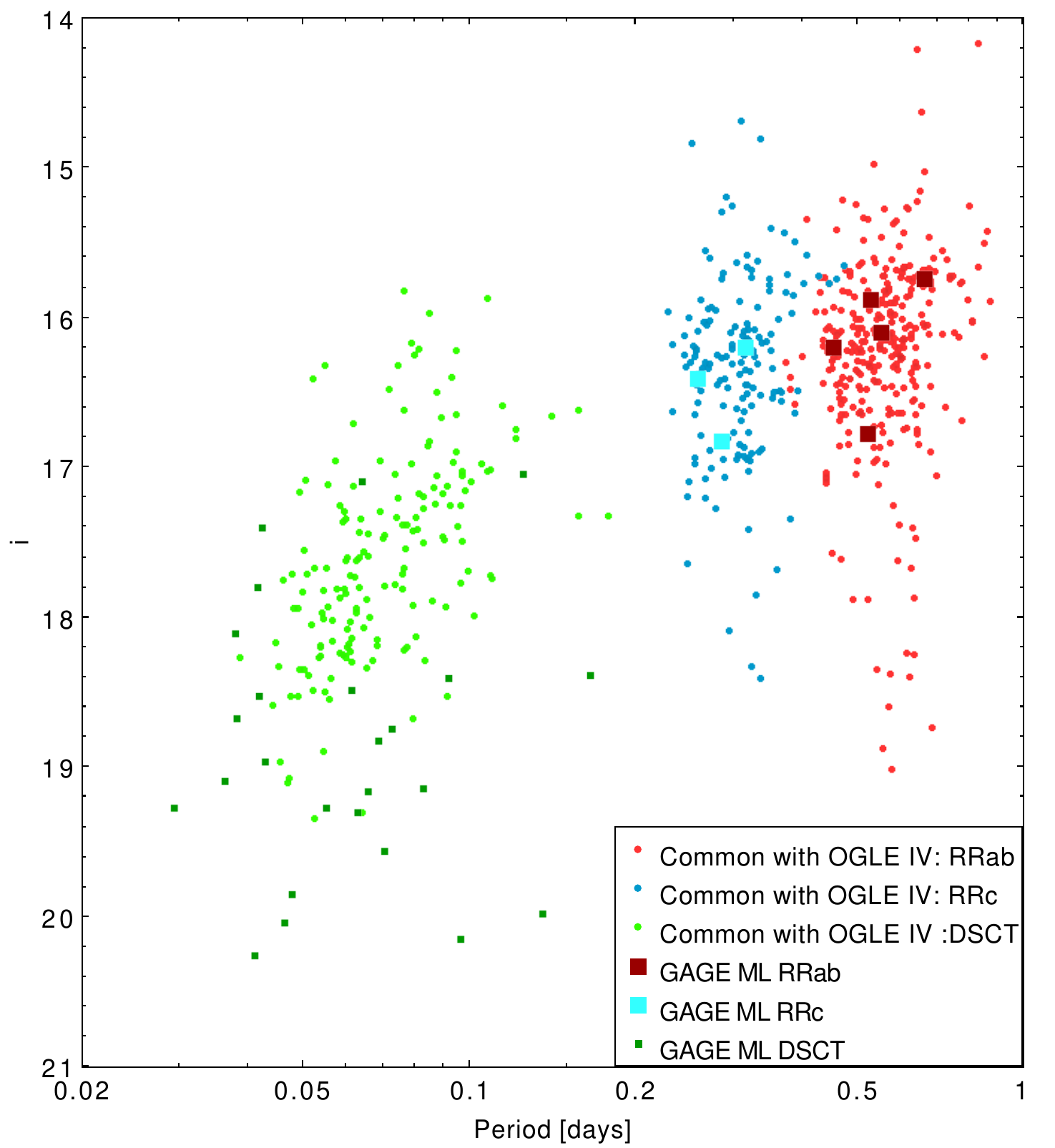}
  \caption{Period--magnitude distributions for the GAGE variable star sample in the $SDSS$ $r$ band  \text{(panel a)} and \text{}$SDSS$ $i$--band (panel b), colour-coded as in Fig.~\ref{fig:bailey}.}
  \label{fig:mag_period_combined}
\end{figure*}

The refined DSCT sample consists of 101 objects, of which 78 meet our high-confidence threshold ($QI \geq 50\%$; dark green squares in Figs.~\ref{fig:bailey} and \ref{fig:mag_period_combined}). These newly identified DSCT candidates predominantly populate a faint regime ($19.0 < r < 21.0$\,mag and $18.5 < i < 20.5$\,mag), extending roughly 1.5--2.0\,mag deeper than the brighter OGLE\,IV baseline ($16.5 < r < 18.5$\,mag and $16.0 < i < 18.5$\,mag).

\subsection{Multi-band expansion of the OGLE\,IV catalogue}
\label{sec:ogle_comparison}
The OGLE\,IV survey provides a foundational, decades-long baseline that underpins modern variable star research across the high-density populations of the Galactic bulge. While this survey offers exquisite temporal sampling primarily in the $I$--band, extending these stellar profiles into the $SDSS$ $ugriz$ filter system offers a critical pathway to investigate wavelength-dependent pulsations. Providing uniform $u$--, $g$--, $r$--, $i$--, and $z$--band measurements for these variables is essential for refining their characterisation within such an interesting environment. By utilising the cross-matched and filtered sample established in Sect.~\ref{sec:method}, which recovers 97.85\% of the known variables (11,897 out of 12,158 stars) within our footprint and in which 72.46\% of the final GAGE sample is matched to OGLE\,IV (4,568 out of 6,304 stars), this work extends the temporal coverage of these inner Galaxy variables. 

Consequently, we provide the first deep, homogeneous $ugriz$ characterisation for this population, bridging long-term OGLE\,IV dataset in the $V$-- and $I$--bands with the next generation wide-field surveys (LSST but also the proposed Wide-Field Spectroscopic Telescope; \citealt{Mainieri2024}). A structural outline of this expanded dataset is detailed in Table~\ref{tab:ogle_crossmatch}, which showcases our derived $ugriz$ calibrated mean magnitudes, phase amplitudes, and associated uncertainties alongside the original OGLE\,IV catalogue metrics.

\section{Summary and conclusions} \label{sec:conclusions}

We have presented the initial results and structural catalogue from a long-term multi-band ($ugriz$) photometric survey targeting a low-reddening window to the highly crowded and differentially obscured MW bulge (the Baade's window B2 field). By utilising the large collecting area of the 4-m Blanco telescope alongside DECam's wide field of view, the work presented in this paper mitigates severe stellar blending through enhanced survey depth and instrumental sensitivity. Furthermore, by extending the time-domain coverage into the redder $z$ passband, our pipeline effectively bypasses the high dust extinction barriers that limit purely optical wide-field surveys in the inner Galaxy. The primary achievements of this study are summarised below:

\begin{enumerate}
    \item A robust high-yield time-domain method: We have developed a reproducible multi-step method within the framework of \text{The GalActic bulGe with plEiadi} (GAGE) project. Processing 461 single-epoch DECam multi-band exposures spanning six years of observations yielded a unified master database containing 30,415,150 distinct stellar sources.
    
    \item Catalogue purity and method validation: By mapping statistical metrics derived from the GLS periodogram and robust RANSAC modelling against a benchmark OGLE\,IV reference population, we established empirical selection boundaries of the GLS power spectrum. Out of our total sources within the DECam B2 footprint, our automated preliminary thresholds isolated 6,914,581 variable candidates. This sample was subsequently streamlined to 3,820,039 sources following the GLS filtering stage. By enforcing our strict 15th percentile high-quality quantile, the candidate pool was systematically reduced to 613,341 high-significance objects. Ultimately, adopting an automatic flagging system of the quality of the fitting of the phased light curves, corroborated by a final visual inspection, allowed our framework to isolate a definitive catalogue of 6,304 high-quality variable star candidates.
    
    \item Supervised ML classification: We implemented an optimised random forest classifier utilising a 25--dimensional feature space constructed from their photometric and Fourier-based properties. Evaluating the global Gini importance scores revealed that while the primary pulsation period remains the dominant anchor ($\sim$29\% of total classification leverage), calibrated redder passbands provide a critical, highly stable secondary decision tier capable of mitigating heavy differential extinction across the inner bulge footprint.

    \item Multi-band expansion of the OGLE\,IV catalogue: We provide a comprehensive multi-band ($ugriz$) wavelength extension for the variables shared with the OGLE\,IV baseline. This dataset complements long-term optical tracking with deep multi-epoch colour and phase profiles, bridging the gap between legacy database and wide-field multi-filter photometric surveys to enable precise wavelength-dependent pulsation analyses in heavily obscured fields. Our catalogue is composed of $4,568$ (72.46\%) already known OGLE\,IV variables that passed our selection criteria, while it uncovered 1,736 ($\sim$ 28\%) non-OGLE\,IV candidates.
    
    \item Substantial expansion of the inner galaxy inventory: Among these non-OGLE\,IV candidates, 198 and 230 are found in the \textit{Gaia}~DR3 and VVV catalogues, respectively. These non-OGLE\,IV sources were classified utilising a random forest ML architecture, resulting in the identification of 15 RRLs (eight RRab and seven RRc stars), 1,555 ECLs, and 101 short-period DSCT variables. 

    \item Benchmark dataset for LSST: The LSST will provide the most complete, deep, and extensive multi-band ($ugrizY$) time-domain mapping of the southern sky. However, given the LSST observational cadence, the robust characterisation of new periodic variables, including precise measurements of periods, mean magnitudes, amplitudes, and morphological classifications, will likely only become feasible starting from the second year of operations. Therefore, it would be of possible general interest to have a reference catalogue of periodic variable stars to compare with the LSST products \citep[see e.g.][]{Ngeow26}. In addition, we provide a large ($\sim 600$k stars) catalogue of suspected variables, which might be of interest for other studies and for comparisons with the LSST alert system.    
\end{enumerate}

The depth and wavelength coverage of the GAGE catalogue demonstrate that high-cadence long-term archival collections can resolve faint variable populations within highly crowded and extinct fields. The resulting multi-band multi-epoch catalogue provides a uniform empirical dataset that is vital for further studies. 

\section*{Data availability}
Tables~\ref{tab:variables} and~\ref{tab:ogle_crossmatch} are only available in electronic form at the CDS via anonymous ftp to cdsarc.u-strasbg.fr (130.79.128.5) or via \url{http://cdsweb.u-strasbg.fr/cgi-bin/qcat?J/A+A/}.

\begin{acknowledgements}
We sincerely thank the anonymous referee for their insightful comments and suggestions, which substantially enhanced the quality of this manuscript. The authors are highly thankful to Abhijit Saha and Elena Pancino for their valuable support and insightful ideas during the development of this work. We acknowledge support by PLEIADI, INAF -- USC VIII, for the availability of high-performance and high-throughput computing resources. 

M.G. and M.D.O. acknowledge financial support from the INAF Large Grant GAGE (\text{GalActic bulGe with plEiadi}, Ob.Fu. 1.05.23.05.24; PI: M. Dall'Ora). M.G. also acknowledges financial support from the project "WST: the Wide-field Spectroscopic Telescope" (Ob.Fu. 1.05.23.01.04; local PI: P. Schipani). J.O.C acknowledges financial support from the project "Partecipazione LSST - Large Synoptic Survey Telescope" (Ob.Fu. 1.05.03.06; local PI: M. Brescia).

M.T. is supported by the Large Grant INAF 2023 MOVIE (PI: M. Marconi). G.D.S. acknowledges funding from \textit{Gaia} DPAC through INAF and ASI (2025-10-HH.0; PI: M. G. Lattanzi) and from INFN, Naples Section, through the QGSKY and Moonlight2 initiatives. D.M. acknowledges support from Fondecyt Project No. 1220724, and from the Center for Astrophysics and Associated Technologies (CATA) by the ANID BASAL projects ACE210002 and FB210003. M.~Marengo and M.~Jones are supported by the National Science Foundation under Grant No. AST-2407965. C.S.G. acknowledges support from the National Agency for Research and Development (ANID) Scholarship Program, Doctorado Nacional 2025-21251677.
\end{acknowledgements}

\bibliography{sample701}{}
\bibliographystyle{aa}

\clearpage           
\onecolumn
\begin{appendix}
\section{Catalogue layout}

\begin{table*}[ht!]
\centering
\scriptsize                         
\setlength{\tabcolsep}{2pt}          
\renewcommand{\arraystretch}{0.85}    

\begin{minipage}[t]{0.48\textwidth}
\centering
\caption{Data schema and column descriptors for the variable candidate catalogue (extract).}
\label{tab:variables}
\begin{tabular}{r l l}
\hline\hline
\noalign{\vskip 1pt}
Col. & Label & Description \\
\noalign{\vskip 1pt}\hline\noalign{\vskip 1pt}
\multicolumn{3}{l}{\text{Stellar coordinates and ephemerides (6,304 sources)}}\\ \noalign{\vskip 1pt}
1 & \texttt{ID\_GAGE} & Unique survey identifier (formatted as \texttt{GAGE\_ccdnumber\_starID}) \\
2 & \texttt{RA} & Right ascension in decimal degrees (J2000) \\
3 & \texttt{DEC} & Declination in decimal degrees (J2000) \\
4 & \texttt{Period} & Primary pulsation or orbital period derived via GLS (d) \\
5 & \texttt{class\_ML} & Primary ML classification label \\
6 & \texttt{$QI$} & ML quality index ($0.0$--$1.0$) \\
\noalign{\vskip 1pt}\hline\noalign{\vskip 1pt}
\multicolumn{3}{l}{\text{DECam multi-band photometric parameters ($ugriz$)}}\\ \noalign{\vskip 1pt}
7 & \texttt{m\_u} & Mean $u$--band magnitude \\
8 & \texttt{e\_m\_u} & Magnitude error ($u$--band) \\
9 & \texttt{A\_u} & Peak-to-peak $u$--band amplitude \\
10 & \texttt{e\_A\_u} & Amplitude error ($u$--band) \\
11 & \texttt{m\_g} & Mean $g$--band magnitude \\
12 & \texttt{e\_m\_g} & Magnitude error ($g$--band) \\
13 & \texttt{A\_g} & Peak-to-peak $g$--band amplitude \\
14 & \texttt{e\_A\_g} & Amplitude error ($g$--band) \\
15 & \texttt{m\_r} & Mean $r$--band magnitude \\
16 & \texttt{e\_m\_r} & Magnitude error ($r$--band) \\
17 & \texttt{A\_r} & Peak-to-peak $r$--band amplitude \\
18 & \texttt{e\_A\_r} & Amplitude error ($r$--band) \\
19 & \texttt{m\_i} & Mean $i$--band magnitude \\
20 & \texttt{e\_m\_i} & Magnitude error ($i$--band) \\
21 & \texttt{A\_i} & Peak-to-peak $i$--band amplitude \\
22 & \texttt{e\_A\_i} & Amplitude error ($i$--band) \\
23 & \texttt{m\_z} & Mean $z$--band magnitude \\
24 & \texttt{e\_m\_z} & Magnitude error ($z$--band) \\
25 & \texttt{A\_z} & Peak-to-peak $z$--band amplitude \\
26 & \texttt{e\_A\_z} & Amplitude error ($z$--band) \\
\noalign{\vskip 1pt}\hline\noalign{\vskip 1pt}
\multicolumn{3}{l}{\text{Light curve Fourier decomposition parameters ($r$--band)}}\\ \noalign{\vskip 1pt}
27 & \texttt{R21} & Fourier amplitude ratio $R_{21} = A_2/A_1$ used in ML classification \\
28 & \texttt{R31} & Fourier amplitude ratio $R_{31} = A_3/A_1$ used in ML classification \\
29 & \texttt{phi21} & Fourier phase parameter $\phi_{21} = \phi_2 - 2\phi_1$ used in ML classification \\
30 & \texttt{phi31} & Fourier phase parameter $\phi_{31} = \phi_3 - 3\phi_1$ used in ML classification\\
\noalign{\vskip 1pt}\hline\noalign{\vskip 1pt}
\multicolumn{3}{l}{\text{External cross-matches and refined classifications for 1,736 non-OGLE\,IV variables}}\\ \noalign{\vskip 1pt}
31 & \texttt{ID\_Gaia} & Source identifier in \textit{Gaia}~DR3 \\
32 & \texttt{Gaia\_period} & Variable period in \textit{Gaia}~DR3 (d) \\
33 & \texttt{class\_Gaia} & Variable classification in \textit{Gaia}~DR3 \\
34 & \texttt{ID\_VVV} & Source identifier in VVV \\
35 & \texttt{VVV\_period} & Variable period in VVV (d) \\
36 & \texttt{class\_VVV} & Variable classification in VVV \\
37 & \texttt{class\_GAGE} & Final adopted GAGE classification \\
38 & \texttt{class\_source} & Classification source (\texttt{ML}, \texttt{Gaia\_LPV}, \texttt{CMD\_LPV}) \\
\noalign{\vskip 1pt}\hline
\end{tabular}
\tablefoot{
Data schema for the comprehensive multi-band photometric catalogue of variable candidates detected within the DECam B2 footprint. This table is available in its entirety in machine-readable format at the CDS.
}
\end{minipage}
\hfill
\begin{minipage}[t]{0.48\textwidth}
\centering
\caption{Data schema and column descriptors for the OGLE\,IV expansion catalogue (extract).}
\label{tab:ogle_crossmatch}
\begin{tabular}{r l l}
\hline\hline
\noalign{\vskip 1pt}
Col. & Label & Description \\
\noalign{\vskip 1pt}\hline\noalign{\vskip 1pt}
\multicolumn{3}{l}{\text{Stellar cross-identifications and ephemerides (4,568 sources)}}\\ \noalign{\vskip 1pt}
1 & \texttt{ID\_GAGE} & Unique internal survey identifier \\
2 & \texttt{ID\_OGLE} & Corresponding OGLE\,IV identifier \\
3 & \texttt{RA} & Right ascension in decimal degrees (J2000) \\
4 & \texttt{DEC} & Declination in decimal degrees (J2000) \\
5 & \texttt{Period} & Derived pulsation/orbital period (d) \\
6 & \texttt{Type} & Primary OGLE\,IV variability classification \\
7 & \texttt{Subtype} & Assigned OGLE\,IV sub-classification type \\
\noalign{\vskip 1pt}\hline\noalign{\vskip 1pt}
\multicolumn{3}{l}{\text{Legacy baseline photometric parameters (OGLE)}}\\ \noalign{\vskip 1pt}
8 & \texttt{m\_I} & OGLE mean $I$--band magnitude \\
9 & \texttt{A\_I} & OGLE peak-to-peak $I$--band amplitude \\
10 & \texttt{m\_V} & OGLE mean $V$--band magnitude \\
11 & \texttt{A\_V} & OGLE peak-to-peak $V$--band amplitude \\
\noalign{\vskip 1pt}\hline\noalign{\vskip 1pt}
\multicolumn{3}{l}{\text{DECam multi-band photometric parameters ($ugriz$)}}\\ \noalign{\vskip 1pt}
12 & \texttt{m\_u} & Mean $u$--band magnitude \\
13 & \texttt{e\_m\_u} & Magnitude error ($u$--band) \\
14 & \texttt{A\_u} & Peak-to-peak $u$--band amplitude \\
15 & \texttt{e\_A\_u} & Amplitude error ($u$--band) \\
16 & \texttt{m\_g} & Mean $g$--band magnitude \\
17 & \texttt{e\_m\_g} & Magnitude error ($g$--band) \\
18 & \texttt{A\_g} & Peak-to-peak $g$--band amplitude \\
19 & \texttt{e\_A\_g} & Amplitude error ($g$--band) \\
20 & \texttt{m\_r} & Mean $r$--band magnitude \\
21 & \texttt{e\_m\_r} & Magnitude error ($r$--band) \\
22 & \texttt{A\_r} & Peak-to-peak $r$--band amplitude \\
23 & \texttt{e\_A\_r} & Amplitude error ($r$--band) \\
24 & \texttt{m\_i} & Mean $i$--band magnitude \\
25 & \texttt{e\_m\_i} & Magnitude error ($i$--band) \\
26 & \texttt{A\_i} & Peak-to-peak $i$--band amplitude \\
27 & \texttt{e\_A\_i} & Amplitude error ($i$--band) \\
28 & \texttt{m\_z} & Mean $z$--band magnitude \\
29 & \texttt{e\_m\_z} & Magnitude error ($z$--band) \\
30 & \texttt{A\_z} & Peak-to-peak $z$--band amplitude \\
31 & \texttt{e\_A\_z} & Amplitude error ($z$--band) \\
\noalign{\vskip 1pt}\hline
\end{tabular}
\tablefoot{
Data schema for the multi-band expansion catalogue of OGLE\,IV variables within the DECam B2 footprint. This table is available in its entirety in machine-readable format at the CDS.
}
\end{minipage}

\end{table*}

\end{appendix}
\end{document}